\documentclass[aps,prd,twocolumn,superscriptaddress,nofootinbib]{revtex4-2}
\pdfoutput=1
\usepackage{graphicx} 
\usepackage{csquotes}
\usepackage{amsmath}
\usepackage{mathtools}
\usepackage{amssymb}  
\usepackage{bm}
\usepackage{float}
\usepackage{subfig}
\renewcommand{\arraystretch}{0.94}
\usepackage[utf8]{inputenc}
\usepackage{xcolor}
\definecolor{darkpurple}{RGB}{90,0,130}
\usepackage[colorlinks=true, citecolor=magenta, linkcolor=blue, urlcolor=darkpurple]{hyperref}
\begin{document}

\title{Electromagnetic Emission from a Black Hole Evaporating  in  External Magnetic Field}

\author{Soumya Samrat Mandal}
\email{mandal15@purdue.edu}
\affiliation{Department of Physics and Astronomy, Purdue University, West Lafayette, Indiana-47907, USA}

\author{Maxim Lyutikov}
\email{lyutikov@purdue.edu}
\affiliation{Department of Physics and Astronomy, Purdue University, West Lafayette, Indiana-47907, USA}

\begin{abstract}
We describe a  classical (non-quantum) radiation process: additional (to Hawking) emission by a black hole evaporating in an external magnetic field in vacuum. The electromagnetic radiation process is completely electric charge-free and bears some resemblance to the Gertsenshtein-Zel'dovich effect. The time evolution of the spacetime metric perturbs a static background magnetic field, inducing a radiative field that acts as an effective electromagnetic source even in the absence of physical charges or currents. To isolate the dynamic effects of the time-dependent spacetime on the external magnetic field, we approximate Hawking radiation as a spherically symmetric outflow of null fluid governed by the prescribed time-dependence of the central mass $M(t)$. 

We employ Laplace transform, which selects the retarded outgoing branch of the electromagnetic response, producing fields proportional to $\Theta(t-r)$ and thereby fixing a causal radiative arrow of emission.
The emitted spectral energy is red-dominated, scaling as $d\mathcal{E}_B/d\omega \propto B_0^2 m_0^2 \tau_H^{-2/3} \omega^{-8/3}$, where $\tau_H$ is the Hawking evaporation timescale, and corresponds to a pure Transverse Electric (TE) mode. There is no final bright burst at the end of the evaporation.  We offer classical analogues for this mechanism as transmission-line emission and, separately, as the displacement current emission from a medium with time-varying dielectric permittivity.
\end{abstract}

\maketitle

\section{Introduction}

Describing the gravitational field and spacetime geometry of a system with a time-dependent mass remains a central challenge in general relativity. While exact solutions like the Vaidya metric \cite{Vaidya,Vaidya2,Vaidya3,coudray} and the Oppenheimer-Snyder collapse model \cite{OS1} provide foundational insights into radial energy flux and gravitational collapse, their highly symmetric nature fails to capture semiclassical radiative effects or produce nontrivial electromagnetic sources when coupled to external fields\footnote{See Appendices \ref{Vaidya} and \ref{Oppenheimer} for details on the Vaidya and Oppenheimer-Snyder spacetimes.}. Instead, evaporating black holes are better modeled using semiclassical frameworks, initiated by Bardeen \cite{Bardeen}, where the classical geometry evolves in response to quantum Hawking emission \cite{Hiscock,Hiscock2,Abdolrahimi}. In this regime, the mass decays as $\dot{M} \propto -1/M^2$ \cite{Hawking2}, generating a non-stationary extension of the Schwarzschild metric.

The timescale of this mass evolution, $\tau_{\rm H}$, fundamentally dictates the electromagnetic response. Because the geometry evolves slowly, it induces electromagnetic modes with exceptionally long wavelengths, $\lambda \sim \tau_{\rm H}$. This macroscopic spatial scale for radiation formation blurs the traditional distinction between near-field and radiative zones. Consequently, a purely quasi-static approximation is inadequate; capturing the full physical picture requires a rigorous analysis of how time-dependent gravitational backgrounds dynamically generate propagating modes. In this work, we investigate how this time-dependent gravitational background dynamically couples to an external, constant magnetic field. Working in the weak-field (test-field) regime where the magnetic field does not backreact on the underlying geometry \cite{Galtsov,Wald74}, we allow for a perturbative treatment of Maxwell’s equations in curved spacetime \cite{WaldB,Poisson}. We treat the black hole mass as a time-dependent function $M(t)$ driven by a spherically symmetric, outgoing massless flux.

By decomposing the electromagnetic field into a background and a radiative perturbation \cite{Bicak1,Bicak2,Bicak3,ReggeWheeler,Zerilli,Teukolsky,Palenzuela,Mosta}, we show that the evolving spacetime forces the otherwise static background magnetic field to become dynamically evolving. Consequently, the background field no longer satisfies the vacuum Maxwell equations ($\nabla_\mu F^{\mu\nu}=0$) independently. Instead, the metric's time dependence couples to the electromagnetic field tensor, acting as an effective source term that drives the emission of purely classical electromagnetic radiation, even in the absence of physical charges or currents.

Given the prevalence of strong magnetic fields in astrophysical environments—such as those surrounding neutron stars, magnetars, and supermassive black holes \cite{Rueda,Hussain,Hooper,Babar}—this mechanism represents a novel, first-order classical correction to the standard quantum Hawking evaporation process \cite{Hawking3,Wald75}. The resulting energy loss channel depends strictly on the background field strength and the initial black hole mass. Beyond evaporating black holes, this formalism establishes a broader framework for studying how gravitational dynamics can source electromagnetic emission in other non-rotating, time-dependent systems, such as primordial black hole formation or spherically symmetric accretion.

\section{Setup for non-rotating, spherically symmetric, time-dependent spacetimes}

In this work, we consider a non-rotating, spherically symmetric Schwarzschild black hole undergoing evaporation via Hawking radiation, in the presence of an external electromagnetic field. In curved spacetime, Maxwell’s equations can be written in covariant form as
\begin{equation}
\label{P1}
\partial_\mu\left(\sqrt{-g}\,F^{\mu\nu}\right)=J^\nu.
\end{equation}

In the absence of a true physical charge or current density, $J^\nu = 0$, and the full electromagnetic field satisfies the vacuum Maxwell's equations: 
\begin{equation}
\label{P2}
\partial_\mu\left(\sqrt{-g}\,F^{\mu\nu}\right)=0.
\end{equation}

In a time-dependent spacetime, it is advantageous to decompose the electromagnetic field into a background component and a dynamical perturbation:
\begin{equation}
\label{P3}
A_\mu = A_\mu^{(0)} + \delta A_\mu,
\end{equation}
where $A^{(0)}_\mu$ is the background vector potential and $\delta A_\mu$ is the radiative or wave-like correction generated by the evolving geometry. Although the full field satisfies Eq.~\eqref{P2}, the background field generally does not satisfy these equations independently when the metric evolves in time. Consequently, the background field acts as an effective source for the perturbations $\delta A_\mu$. 

\subsection{Electromagnetic emission without a source}

Before computing the effective source term, the background field $A_\mu^{(0)}$ must be defined. In the weak-field limit, where the metric is $g_{\mu\nu}=\eta_{\mu\nu}+h_{\mu\nu}$, this decomposition can be approached in two equivalent ways. 

One approach is to retain the flat-space uniform magnetic field as the background, treating all gravitational effects—both static and dynamical—as perturbations. In this case, the background field does not satisfy the curved-spacetime Maxwell equations independently, and the entire gravitational effect is encoded in the effective source term. Alternatively, one may absorb the static part of $h_{\mu\nu}$ into the background configuration, ensuring $A_\mu^{(0)}$ satisfies Maxwell’s equations in the static curved spacetime:
\begin{equation}
\partial_\mu\left(\sqrt{-g_{\text{static}}}\,F^{(0)\mu\nu}\right)=0,
\end{equation}
leaving only the explicitly time-dependent part of $h_{\mu\nu}$ to drive radiation. Both approaches are equivalent up to the relevant perturbative order.

We adopt the latter approach. The background magnetic field is defined as the static configuration corresponding to the Schwarzschild spacetime, meaning only the time-dependent part of the metric contributes to the effective source term. From Eq.~\eqref{P3}, the background electromagnetic tensor is:
\begin{equation}
\label{P4}
F_{\mu\nu}^{(0)}=\partial_\mu A_\nu^{(0)}-\partial_\nu A_\mu^{(0)}.
\end{equation}
The full electromagnetic tensor is then:
\begin{equation}
\label{P5}
F_{\mu\nu}=F_{\mu\nu}^{(0)}+\delta F_{\mu\nu}.
\end{equation}
Substituting this into the vacuum Maxwell equations yields:
\begin{equation}
\label{P6}
\begin{split}
&\partial_\mu\left[\sqrt{-g}\,\left(F^{(0)\mu\nu}+\delta F^{\mu\nu}\right)\right]=0\\
&\Rightarrow\partial_\mu(\sqrt{-g}\,\delta F^{\mu\nu})=S^\nu_{\text{eff}},
\end{split}
\end{equation}
where the effective source term is defined as:
\begin{equation}
\label{P7}
S^{\nu}_{\text{eff}}=-\partial_\mu(\sqrt{-g}\,F^{(0)\mu\nu}).
\end{equation}

Thus, when the spacetime becomes time-dependent, a non-vanishing effective current is generated. This current does not correspond to a physical charge; rather, it is a macroscopic consequence of the time-dependent metric acting on the background magnetic field, driving electromagnetic radiation.

To evaluate Eq.~\eqref{P7}, we require the static background field and the evolving metric. Assuming an axisymmetric background magnetic field aligned along the spatial $z$-direction, the vector potential has only a non-vanishing azimuthal component, $A_\phi^{(0)}$. In the static weak-field limit, this is given in isotropic Schwarzschild coordinates $(t,r',\theta,\phi)$ by:
\begin{equation}
\label{(2)}
A_{\phi}^{(0)}(r',\theta)= \frac{B_0 r'^2 \sin^2\theta}{2},
\end{equation}
where $B_0$ denotes the background magnetic field strength. This expression provides the static background configuration upon which the effects of the time-dependent spacetime evolution are treated as perturbations.

\subsection{The spacetime of a time-evolving central mass: spherically symmetric, time-dependent metric}

A full semiclassical description of an evaporating black hole is highly nontrivial, requiring solutions to Einstein's equations coupled to the quantum stress-energy tensor. Because our primary focus is the macroscopic electromagnetic response rather than microscopic backreaction, we adopt a broader, simplified approach. We model the background using a prescribed, time-dependent effective mass, capturing the essential gravitational dynamics driving the electromagnetic emission.

We begin with a general spherically symmetric, time-dependent metric motivated by \cite{Bardeen}:
\begin{equation}
\label{(3)i)}
\begin{split}
ds^2 &= -\,e^{2\Phi(r',t)} \left(\frac{2r'-m(r',t)}{2r'+m(r',t)}\right)^2 dt^2 \\
&\quad + \left(1+\frac{m(r',t)}{2r'}\right)^4 \left[ dr'^2+r'^2 d\Omega^2 \right]
\end{split}
\end{equation}
Henceforth, for brevity, we denote the isotropic radial coordinate by $r$ rather than $r'$. The function $m(r,t)$ represents the effective mass contained within a sphere of radius $r$ at time $t$, consisting of the central mass $M(t)$ together with any massless energy propagating outward.

Rather than deriving $\Phi(r,t)$ and $m(r,t)$ from a fully self-consistent quantum stress-energy tensor, we treat the metric as an effective background governed by a prescribed central mass evolution $M(t)$. We fix the unknown metric functions by demanding the vanishing of the angular Einstein tensor components. This strictly enforces a stress-energy tensor with purely radial energy flux and zero tangential pressure. Consequently, Eq.~\eqref{(3)i)} serves as a minimal, physically consistent framework for modeling spherically symmetric radial energy flow (such as Hawking emission or spherical accretion) and evaluating the resulting electromagnetic response.

For notational compactness in the ensuing derivations, we suppress the explicit $(r,t)$ dependence of the functions $m$ and $\Phi$. The non-zero elements of the contravariant Einstein tensor $G^{\mu \nu}= R^{\mu \nu}-\frac{1}{2} R g^{\mu \nu}$ are obtained as follows:
\begin{equation}
\label{Einstein Generalized}
\setlength{\jot}{0pt}
\begin{aligned}
G^{tt} &= \frac{4 e^{-4 \Phi}}{(2r - m)^4 (2r + m)^3} \Big[ 3 (2r + m)^5 \dot{m}^2 \\
&\quad - 16 e^{2 \Phi} r^4 (2r - m)^2 m'' \Big], \\
G^{tr} &= G^{rt} = \frac{64 e^{-2 \Phi} r^4}{(4r^2 - m^2)^3} \Big[ -\dot{m} \big( 2 - m' \\
&\quad - (2r - m) \Phi' \big) + (2r - m) \dot{m}' \Big], \\
G^{rr} &= \frac{64 e^{-2 \Phi} r^4}{(2r - m)^3 (2r + m)^9} \Big[ -2 (4r - m) (2r + m)^5 \dot{m}^2 \\
&\quad + (2r - m) (2r + m)^6 \dot{m} \dot{\Phi} + (2r - m) f_1 \Big], \\
G^{\theta\theta} &= \frac{64 e^{-2\Phi} r^2}{(2r - m)^3 (2r + m)^9} \Big[ -2 (4r - m) (2r + m)^5 \dot{m}^2 \\
&\quad + (2r - m) (2r + m)^6 \dot{m} \dot{\Phi} + (2r - m) f_2 \Big], \\
G^{\phi\phi} &= G^{\theta\theta}/\sin^2\theta
\end{aligned}
\end{equation}
where dots and primes denote derivatives with respect to $t$ and $r$, respectively, and the functions $f_1$ and $f_2$ are:
\begin{equation*}
\setlength{\jot}{0pt}
\begin{aligned}
f_1 &= - (2r + m)^6 \ddot{m} + 8 e^{2 \Phi} r^3 (2r - m) \Big[ 2 m' (m - r m') \\
&\quad - (2r - m) \big[ m - 2r(1 + m') \big] \Phi' \Big], \\
f_2 &= - (2r + m)^6 \ddot{m} + 4 e^{2\Phi} r^3 (2r - m) \Big[ -2 m \big[ m' \\
&\quad - r(4\Phi' - m'') \big] - m^2 \big( \Phi' + r \Phi'^2 + r \Phi'' \big) \\
&\quad + 2r \big( m'^2 - 4r m' \Phi' \big) + 2r \big( \Phi' + r \Phi'^2 + r \Phi'' \big) \Big].
\end{aligned}
\end{equation*}

In the present scenario, we restrict our attention to spacetimes governed solely by the changing central mass $M(t)$. To reduce the metric to a single dynamical degree of freedom, $\Phi(r,t)$ and $m(r,t)$ must be functionally related. We therefore impose:
\begin{equation}
\label{(3)ii)}
\Phi(r,t)=\Phi\big(m(r,t)\big)
\end{equation}

\begin{itemize}
    \item \textbf{Special Case: $\Phi\big(m(r,t)\big)$ as a linear function of $m(r,t)$}
\end{itemize}

Working in the weak-field regime, we can consistently expand the Einstein tensor, retaining only terms linear in $m(r,t)$. Within this approximation, we define a linear functional dependence:
\begin{equation}
\label{(3)iii)}
\Phi\big(m(r,t)\big)=\frac{m(r,t)}{r_0},
\end{equation}
where $r_0$ is a constant length scale characterizing the system to be determined. Keeping terms up to $\mathcal{O}(m)$, the metric in \eqref{(3)i)} reduces to:
\begin{equation}
\label{4}
\begin{split}
g_{tt} &= -1 + \frac{2\, m(r,t)}{r} - \frac{2 \,m(r,t)}{r_0} \\
g_{rr} &= 1 + \frac{2\, m(r,t)}{r} \\
g_{\theta \theta} &= r^2\,g_{rr} \\
g_{\phi \phi} &= r^2 \sin^2\theta\,g_{rr}
\end{split}
\end{equation}

In the complete weak-field limit, the static Schwarzschild geometry introduces $\mathcal{O}(m)$ curvature corrections across all metric components, which subsequently dress the background magnetic field. However, these purely static transverse distortions do not contribute to the leading-order time-dependent radiation. To rigorously isolate the dynamic electromagnetic response, we retain the $\mathcal{O}(m)$ gravitational corrections exclusively in the temporal and radial sectors ($g_{tt}$ and $g_{rr}$), as these couple directly to the radial energy flow. The transverse angular components are truncated at zeroth order ($g_{\theta\theta} \approx r^2$, $g_{\phi\phi} \approx r^2\sin^2\theta$). This mathematically filters out the static gravitational dressing, ensuring the effective source is generated entirely by the spacetime dynamics.

We further assume the mass evolution is slow ($\dot{m}, m' \ll 1$), allowing us to drop quadratic and higher-order derivatives. The leading-order Einstein tensor then takes the simplified form:
\begin{equation}
\label{5}
\begingroup
\setlength{\arraycolsep}{3pt}
\renewcommand{\arraystretch}{0.82}
G^{\mu \nu}=
\left[
\begin{array}{cccc}
-\frac{ 2\, r m''}{r^2} & -\frac{2(\dot{m} - r \dot{m}')}{r^2} & 0 & 0 \\
-\frac{2(\dot{m} - r \dot{m}')}{r^2} & -\frac{2(r_0 \ddot{m} - m')}{r\, r_0} & 0 & 0 \\
0 & 0 & \frac{F[m]}{r^3\,r_0} & 0 \\
0 & 0 & 0 & \frac{F[m]}{r^3\, r_0\sin^2\theta}
\end{array}
\right]
\endgroup
\end{equation}
where the functional $F[m]$ is given by:
\begin{equation}
\label{6}
F[m]=r\, m'' + m' -2 r_0\, \ddot{m}
\end{equation}

By construction, this tensor satisfies local energy-momentum conservation ($\nabla_\mu G^{\mu \nu}=0$). Enforcing our prior physically motivated constraint of vanishing transverse pressure ($G^{\theta\theta}=G^{\phi\phi}=0$) forces $F[m]=0$, which reduces \eqref{6} to:
\begin{equation}
\label{10}
r\, m''(r,t) + m'(r,t) - 2 r_0\, \ddot{m}(r,t)=0.
\end{equation}
Equation \eqref{10} serves as the governing equation for the mass function $m(r,t)$, alongside the to-be-determined scale $r_0$. 

\subsection{Effective source term for Maxwell's equations}
\label{source}

With the static vacuum solution for $A_\phi^{(0)}(r,\theta)$ in the weak-field limit given by \eqref{(2)}, the source term for Maxwell's equations in the spacetime around a black hole with a time-varying mass is given by:
\begin{equation}
\label{27}
S^\phi=\partial_r\left(\sqrt{-g}g^{rr}g^{\phi \phi}F_{r \phi}^{(0)}\right)+\partial_\theta\left(\sqrt{-g}g^{\theta\theta}g^{\phi \phi}F_{\theta \phi}^{(0)}\right)
\end{equation}
where $F^{(0)}_{r\phi}=B_0\,r\sin^2\theta$ and $F^{(0)}_{\theta\phi}=B_0\,r^2\sin\theta\cos\theta$. Suppressing the explicit $(r,t)$ dependence for brevity, the inverse metric in the small $m(r,t)$ limit is obtained from \eqref{4} as follows:
\begin{equation}
\label{28}
g^{\mu \nu}=\begin{bmatrix}
-1 - \frac{2m}{r} + \frac{2m}{r_0} & 0 & 0 & 0 \\[1.5ex]
0 & 1 - \frac{2m}{r} & 0 & 0 \\[1.5ex]
0 & 0 & \frac{1}{r^2} & 0 \\[1.5ex]
0 & 0 & 0 & \frac{1}{r^2\,\sin^2\theta}
\end{bmatrix}
\end{equation}
Substituting the inverse metric elements from \eqref{28} into \eqref{27}, we obtain to leading order:
\begin{equation}
\label{28b}
S^\phi=-2\,B_0\,m'(r,t)\sin\theta    
\end{equation}

The appearance of $m'(r,t)$ admits a clear physical interpretation: the time-dependent central mass $M(t)$ induces a radial redistribution of energy. As energy propagates outward, the effective mass function $m(r,t)$ develops a nontrivial radial profile. Its spatial derivative explicitly encodes this outward energy flux. Consequently, the emitted electromagnetic radiation is directly sourced by the radial transport of energy driven by the dynamically evolving gravitational field.

With the general source term established, the construction is applicable to any prescribed spherically symmetric, time-dependent effective mass profile $m(r,t)$. We now treat Hawking evaporation as the main application. Additional auxiliary cases, involving oscillating and exponentially decaying mass profiles, are deferred to Appendices~\ref{oscillate} and~\ref{Laplace}, where they serve as consistency checks of the same construction.

\section{Particular case for an evaporating blackhole emitting Hawking Radiation}
\label{Hawking}

We now turn to a black hole evaporating through Hawking radiation. Since the induced source is controlled by $m'(r,t)$, we first construct the causal radial profile $m(r,t)$ associated with the outgoing massless flux.

In this case, the decaying black hole mass $M(t)$ is modeled as
\begin{equation}
\label{18}
M(t)=m_0\left(1-\frac{t}{\tau_{\rm H}}\right)^{1/3}
\end{equation}
where $m_0$ is the initial black hole mass and $\tau_{\rm H}$ is the Hawking evaporation time. The calculation below treats this mass law as prescribed boundary data at the inner surface $r=2m_0$. The geometry is taken to be stationary for $t\leq 0$ and begins to evolve for $t>0$, which makes the problem a retarded initial-value problem.

\subsection{Causal mass profile, boundary conditions, and Laplace-space solution}

As discussed in Sec.~\ref{source}, the spacetime-dependent mass function is determined by Eq.~\eqref{10}. For the evaporating case we separate the stationary part of the geometry from the dynamical mass loss by writing
\begin{equation}
\label{Hmassprofile}
m(r,t)=\begin{cases}
m_0 \quad\quad\quad\,\,\,\,\,\,\,, & t \leq 0 \\
m_0-\tilde{m}(r,t), & t > 0
\end{cases}
\end{equation}
where $\tilde{m}(r,t)$ represents the outgoing massless component generated by evaporation. The prescribed Hawking mass loss is imposed at the inner boundary, so that
\begin{equation}
\label{Hboundary}
m(r=2m_0,t)=\begin{cases}
m_0 \quad\quad\quad\,\,, & t \leq 0 \\
m_0-M(t), & t > 0 .
\end{cases}
\end{equation}
where $\tilde{m}(r=2m_0,t)=M(t)$. This fixes the normalization; Eq.~\eqref{10} then determines the radial propagation of the outgoing flux.

Because the system is stationary before $t=0$ and evolves causally afterward, we use the one-sided Laplace transform
\begin{equation}
\label{Hlaplace}
m(r,s)=\int_0^\infty e^{-st}\, m(r,t) \,dt
\end{equation}
with $\Re(s)>0$. Substituting this transform into Eq.~\eqref{10} gives the radial equation
\begin{equation}
\label{Hradialeq}
r \, m''(r)+ m'(r)-2 r_0 s^2 \, m(r)=0 .
\end{equation}
Its general solution is
\begin{equation}
\label{Hgeneralsolution}  
m(r,s)=c_1(s) \, K_0\left(2 \sqrt{2rr_0} \, s\right) + 
c_2(s) \, I_0\left(2 \sqrt{2rr_0} \, s\right),
\end{equation}
where $I_0$ and $K_0$ denote modified Bessel functions of the first and second kinds. The choice between these two branches is fixed by the physical boundary condition at large $r$.

Since the blackhole is decaying by losing mass, we want the source to emit outgoing null waves of positive energy density in the Fourier domain. In Laplace space, this retarded/outgoing condition is implemented by requiring the solution to remain bounded as $r \to \infty$ for $\Re(s) > 0$, which eliminates the exponentially growing $I_0$ branch of the modified Bessel functions and leaves the decaying $K_0$ solution. This requires $c_2(s)=0$. Further, upon analytic continuation $s \to i\omega + 0^+$, this choice precisely reproduces the outgoing Hankel function $H_0^{(1)}$. Using $K_0(z)=\frac{\pi i}{2}H_0^{(1)}(iz)$, we may write the causal outgoing solution as
\begin{equation}
\label{Houtgoingsolution}
m(r,s)=c_1(s) \, K_0\left(2 \sqrt{2rr_0} \, s\right)=c_1(s)\, H_0^{(1)}\left(2 i\sqrt{2rr_0} \, s\right).
\end{equation}

With $M(t)$ given by \eqref{18}, the Laplace transform of $m(r=2m_0,t)$ from \eqref{Hboundary} is

\begin{equation}
\label{19}
\begin{aligned}
m(r=2m_0,s)=m_0\Bigg[&\frac{1}{s} \\
&- \frac{e^{-s \tau_{\rm H}} \left(-\frac{1}{\tau_{\rm H}}\right)^{\frac{1}{3}} 
\Gamma\left(\frac{4}{3}, -s \tau_{\rm H}\right)}{s^{\frac{4}{3}}}\Bigg]
\end{aligned}
\end{equation}
Comparing this with $m(r,s)$ at $r=2m_0$ from \eqref{Houtgoingsolution}, we obtain

\begin{equation}
\label{20}
\begin{aligned}
c_1(s)=&\frac{m_0}{ H_{0}^{(1)}\left( 4i\sqrt{m_{0}r_0} \, s \right) }
\Bigg[  \frac{1}{s} \\
&-\frac{e^{-s \tau_{\rm H}} 
   \left( -\tfrac{1}{\tau_{\rm H}} \right)^{\tfrac{1}{3}} 
   \Gamma\!\left( \tfrac{4}{3}, -s \tau_{\rm H} \right)}{s^{4/3}}  \Bigg]
\end{aligned}
\end{equation}
Substituting this into \eqref{Houtgoingsolution}, we get

\begin{equation}
\label{21}
\begin{aligned}
m(r,s)=&\frac{m_0\,H_{0}^{(1)}\left( 2i \sqrt{2rr_0} \, s \right)}
{ H_{0}^{(1)}\left( 4i \sqrt{m_{0}r_0} \, s \right) } 
\Bigg[  \frac{1}{s} \\
&-\frac{e^{-s \tau_{\rm H}} 
   \left( -\tfrac{1}{\tau_{\rm H}} \right)^{\tfrac{1}{3}} 
   \Gamma\!\left( \tfrac{4}{3}, -s \tau_{\rm H} \right)}{s^{4/3}}  \Bigg]
\end{aligned}
\end{equation}

We next determine $r_0$, which fixes the characteristic length scale of the evaporation process. Since Hawking evaporation is slow, the mass varies on the long timescale $\tau_{\rm H}$; in Laplace space, this corresponds to the small-$s$ regime, with the dominant contribution coming from modes $s\sim \tau_{\rm H}^{-1}$.
\begin{equation}
\label{22}
m(r,s)\approx-\frac{m_0 \left(-\frac{1}{\tau_{\rm H}}\right)^{1/3}\, \Gamma\left(\frac{4}{3}\right)}{s^{4/3}}
\end{equation}
The leading small-$s$ term is radially independent and therefore carries no information about the propagation scale. Thus, $r_0$ is fixed by matching the first nontrivial $r$-dependent correction:

\begin{equation}
\label{23}
\begin{aligned}
m(r,s)\approx&-\frac{m_0 \left(-\frac{1}{\tau_{\rm H}}\right)^{1/3}\, \Gamma\left(\frac{4}{3}\right)}{s^{4/3}}\\
&-2 m_0\, r\, r_0\, s^{2/3} \left(-\frac{1}{\tau_{\rm H}}\right)^{1/3} \Gamma\left(\frac{4}{3}\right)
\end{aligned}
\end{equation}
For comparison with a retarded Hawking profile, we define
\begin{equation*}
\tilde{\mathcal{L}}(r,t)=-m_0\left(1-\frac{t-r}{\tau_{\rm H}}\right)^{1/3} .
\end{equation*}
Its Laplace transform gives the first radial correction

\begin{equation}
\label{24}
\begin{aligned}
&\int_0^\infty e^{-st}\, m_0\left(1-\frac{t-r}{\tau_{\rm H}}\right)^{1/3} \,dt\\
&\quad=-r \,\tau_{\rm H}\,m_0 \, s^{2/3} \left(-\frac{1}{\tau_{\rm H}}\right)^{1/3} \, \Gamma\left(\frac{4}{3}\right)
\end{aligned}
\end{equation}
For positive $t$, $\tilde{\mathcal{L}}(r,t)$ represents a mass profile of initial magnitude $m_0$ that follows Hawking-like decay at retarded time $t_r=t-r$. This comparison supplies the physical normalization of the radial propagation scale. The leading term alone cannot determine this scale, because it is independent of the radial position. The second term in the small-$s$ expansion of $m(r,s)$ in \eqref{23} is therefore compared with the Laplace transform of $\tilde{\mathcal{L}}(r,t)$. Comparing \eqref{23} and \eqref{24}, we obtain
\begin{equation}
\label{25}
r_0=\frac{\tau_{\rm H}}{2}
\end{equation}
where $\tau_{\rm H}$ is the lifetime of the evaporating Schwarzschild black hole, which in geometrized units is given by
\begin{equation}
\label{26}
\tau_{\rm H}=5120\, \pi\, m_0^3
\end{equation}

With $m(r,s)$ and $r_0$ given by \eqref{21} and \eqref{25}, we now have the complete metric of an evaporating black hole in the weak-field approximation described by \eqref{4}. 

\subsection{Source term for modified Maxwell's equations in the case of Hawking Radiation}
We now compute the corresponding source term for Maxwell's equations arising from the dynamically evolving spacetime surrounding the evaporating black hole. Substituting $r_0=\tau_{\rm H}/2$ into \eqref{21}, we obtain
\begin{equation}
\label{29}
\begin{aligned}
m(r,s)=&
\frac{m_0\,H_{0}^{(1)}\!\left( 2i \sqrt{r\tau_{\rm H}}\, s \right)}
     {H_{0}^{(1)}\!\left( 2i \sqrt{2m_{0}\tau_{\rm H}}\, s \right)}
\\[4pt]
&\times
\left[
\frac{1}{s}
-
\frac{
e^{-s \tau_{\rm H}}
\left( -\tfrac{1}{\tau_{\rm H}} \right)^{1/3}
\Gamma\!\left( \tfrac{4}{3}, -s \tau_{\rm H} \right)
}{
s^{4/3}
}
\right].
\end{aligned}
\end{equation}

Since $r_0=\tau_{\rm H}/2$ sets the characteristic radial scale of variation of the source, the leading contribution to the source term is obtained in the regime where the argument of the Hankel function is small. This requires $\sqrt{r\tau_{\rm H}}\,|s|\ll 1$, which corresponds to the near-zone region $r\ll r_0$ for the relevant low-frequency band $s\tau_{\rm H}\ll 1$. Rather than performing a formal small-argument series expansion, whose normalization is logarithmically sensitive, we extract the dominant dynamical contribution that governs the late-time evolution.\footnote{In the late-time regime, the Bromwich integral for the inverse Laplace transform is overwhelmingly dominated by small Laplace frequencies $s \sim 1/t \ll 1$. Over this narrow dominant band, the logarithmic term in the denominator of the exact Hankel solution varies so slowly that it acts as an $\mathcal{O}(1)$ constant amplitude factor. The dynamical time evolution is therefore strictly governed by the isolated pole structure $\sim (1 - s\tau_{\rm H})/s$, yielding the correct late-time behavior up to logarithmically small corrections to the overall amplitude.}

The radially independent part of \eqref{29} does not contribute to the electromagnetic source. The leading nonzero contribution is instead controlled by the first radial dependence of the numerator Hankel function. Differentiating $m(r,s)$ with respect to $r$ in the combined near-zone and small-$s$ limits gives, to leading order,
\begin{equation}
\label{29b}
m'(r,s)\approx-\frac{m_0 \left(-\frac{1}{\tau_{\rm H}}\right)^{1/3}\, \Gamma\left(\frac{4}{3}\right)}{r\,s^{4/3}}
\end{equation}
Substituting \eqref{29b} into \eqref{28b}, the source term in Laplace space becomes
\begin{equation}
\label{32}
S^\phi(s)=\frac{2B_0\,m_0\sin\theta}{r}\Bigg[\frac{ \left(-\frac{1}{\tau_{\rm H}}\right)^{1/3}\, \Gamma\left(\frac{4}{3}\right)}{s^{4/3}}\Bigg]
\end{equation}

\subsection{Power emitted by an evaporating black hole decaying by emitting Hawking radiation}

Having obtained the source term, we now solve Maxwell's equations from \eqref{P6} in the spacetime surrounding the evaporating black hole. The time-dependent mass profile induces an effective source, rendering the equations inhomogeneous.

We are interested in the radiation observed at large distances, where the spacetime is asymptotically flat and the metric in \eqref{P6} reduces to that of Minkowski space. The effects of the evolving geometry are therefore entirely encoded in the source term $S^\phi$, while the field propagation occurs in flat spacetime.

In this regime, the vector potential acquires a time-dependent perturbation $A_\phi(t,r,\theta)$, which represents the correction to the static background field arising from the evolving spacetime. For convenience, we denote $A_\phi(t)\equiv A_\phi(t,r,\theta)$ and solve the equation in Laplace space. Defining
\begin{equation}
\label{(5)}
A_{\phi}(s)= \int_{0}^{\infty} e^{-s t} A_{\phi}(t) \, dt,
\end{equation}
and using the source term from \eqref{32}, Maxwell's equations reduce to the following inhomogeneous equation:
\begin{equation}
\label{(4)a))}
\begin{aligned}
&\frac{s^2 A_\phi(s)}{\sin\theta}
+\frac{\partial_r^2 A_\phi(s)}{\sin\theta}
+\frac{\partial_\theta^2 A_\phi(s)}{r^2\sin^2\theta}
-\frac{\cos\theta}{r^2\sin^2\theta}\,\partial_\theta A_\phi(s)\\
&=
\frac{2B_0m_0\sin\theta}{r}
\left[
\frac{
\left(-\frac{1}{\tau_{\rm H}}\right)^{1/3}
\Gamma\left(\frac{4}{3}\right)}
{s^{4/3}}
\right].
\end{aligned}
\end{equation}

Taking the ansatz for the covariant component of the vector potential, $A_\phi(s)=f(r,s)\sin^2\theta$, helps in separating the radial part of the equation from the angular part. The simplified radial equation then becomes
\begin{equation}
\label{(7)}
f''(r,s)-\left(\frac{2}{r^2}+s^2\right)f(r,s)
=
\frac{
2B_0m_0
\left(-\frac{1}{\tau_{\rm H}}\right)^{1/3}
\Gamma\left(\frac{4}{3}\right)}
{r\,s^{4/3}}
\end{equation}
where $'$ denotes differentiation with respect to $r$. The solution to \eqref{(7)} is given by
\begin{equation}
\label{(8)}
\begin{aligned}
f(r,s)=&\sqrt{\frac{2}{\pi s}}
\Bigg[
c_1\left(-\cosh(rs)+\frac{\sinh(rs)}{rs}\right)\\
&\quad
+i\,c_2\left(\sinh(rs)-\frac{\cosh(rs)}{rs}\right)
\Bigg]+f_p(r,s).
\end{aligned}
\end{equation}
Here $f_p(r,s)$ is the particular solution to \eqref{(7)} and is given by
\begin{equation}
\label{(9)}
f_p(r,s)=
-\frac{
2B_0m_0
\left(-\frac{1}{\tau_{\rm H}}\right)^{1/3}
\Gamma\left(\frac{4}{3}\right)}
{r\,s^{10/3}}.
\end{equation}

Once we have obtained the form of $f(r,s)$, we convert from the covariant coordinate component $A_\phi$ to the physical azimuthal component in the orthonormal spherical basis. Since the coordinate basis vector in the $\phi$ direction has norm $r\sin\theta$, the normalized physical component is given by
\begin{equation}
\label{9b}
A_{\hat{\phi}}(s)
=
\frac{A_\phi(s)}{\sqrt{g_{\phi\phi}}}
=
f(r,s)\frac{\sin\theta}{r}.
\end{equation}

The constants $c_1$ and $c_2$ can be determined by imposing the appropriate normalization conditions. We want $A_{\hat{\phi}}(s)$ to be regular as $r\to0$. This requires the coefficient of the divergent $1/r^2$ term to vanish as $r\to0$, which gives
\begin{equation}
\label{c2Hawking}
c_2=
\frac{
\sqrt{\pi}\,B_0m_0(1+i)
\left(-\frac{1}{\tau_{\rm H}}\right)^{1/3}
\Gamma\left(\frac{4}{3}\right)}
{s^{11/6}}.
\end{equation}
Further, demanding
\begin{equation*}
A_{\hat{\phi}}(s)\approx \frac{e^{-rs}\sin\theta}{r}
\qquad \text{as } r\to\infty,
\end{equation*}
we get
\begin{equation}
\label{c1Hawking}
c_1=
-\frac{
\sqrt{\pi}\,B_0m_0(1-i)
\left(-\frac{1}{\tau_{\rm H}}\right)^{1/3}
\Gamma\left(\frac{4}{3}\right)}
{s^{11/6}}.
\end{equation}
As a result, in the large-$r$ limit, the vector potential in the orthonormal spherical basis becomes
\begin{equation}
\label{9c}
\mathbf{A}(s)=
\left\{
0,\,
0,\,
\frac{
2B_0m_0e^{-rs}
\left(-\frac{1}{\tau_{\rm H}}\right)^{1/3}
\Gamma\left(\frac{4}{3}\right)\sin\theta}
{r\,s^{7/3}}
\right\}.
\end{equation}

Now we determine the asymptotic form of the magnetic field for large $r$, given by $\mathbf{B}=\boldsymbol{\nabla}\times\mathbf{A}$:
\begin{equation}
\label{(10)}
\mathbf{B}(s)=
\left\{
\mathcal{O}\!\left(\frac{1}{r^2}\right),\,
\frac{
2B_0m_0e^{-rs}
\left(-\frac{1}{\tau_{\rm H}}\right)^{1/3}
\Gamma\left(\frac{4}{3}\right)\sin\theta}
{r\,s^{4/3}},\,
0
\right\}.
\end{equation}
For large $r$, it is safe to drop terms with $1/r^2$ dependence. As a result, only $B_{\hat\theta}(s)$ survives. Similarly, the electric field is obtained as follows:
\begin{equation}
\label{(11)}
\mathbf{E}(s)=
\left\{
0,\,
0,\,
\frac{
2B_0m_0e^{-rs}
\left(-\frac{1}{\tau_{\rm H}}\right)^{1/3}
\Gamma\left(\frac{4}{3}\right)\sin\theta}
{r\,s^{4/3}}
\right\}.
\end{equation}
Equations \eqref{(10)} and \eqref{(11)} give the asymptotic forms of the electric and magnetic fields in Laplace space in geometrized units.

However, we are interested in obtaining an expression for the spectral power emitted in this process. For this, we first perform the inverse Laplace transformation from the Laplace domain $s$ to time $t$, followed by a Fourier transformation from time $t$ to frequency $\omega$.

The inverse Laplace transform needed for the radiative electric and magnetic fields is
\begin{equation}
\label{LTsource}
\mathcal{L}^{-1}\!\left[
\frac{2B_0m_0 e^{-rs}\Gamma(4/3)\sin\theta}{r\,s^{4/3}}
\left(-\frac{1}{\tau_{\rm H}}\right)^{1/3}
\right].
\end{equation}
Equivalently, defining
\begin{equation}
\label{LTsourceF}
F(s)=\frac{2B_0m_0\Gamma(4/3)\sin\theta}{r}
\left(-\frac{1}{\tau_{\rm H}}\right)^{1/3}\frac{e^{-rs}}{s^{4/3}},
\end{equation}
the inverse transform is given by the Bromwich integral
\begin{equation}
\label{LTbromwichDef}
f(t)=\frac{1}{2\pi i}\int_{\gamma-i\infty}^{\gamma+i\infty}e^{st}F(s)\,ds .
\end{equation}
Here the Bromwich contour is taken along $\Re(s)=\gamma$ with $\gamma>0$, so that it lies to the right of the branch point at $s=0$ and within the domain of analyticity of the Laplace transform. Substituting \eqref{LTsourceF}, we obtain
\begin{equation}
\label{LTbromwich}
\begin{aligned}
f(t)=&\frac{2B_0m_0\Gamma(4/3)\sin\theta}{r}
\left(-\frac{1}{\tau_{\rm H}}\right)^{1/3} \\
&\times \frac{1}{2\pi i}
\int_{\gamma-i\infty}^{\gamma+i\infty}
\frac{e^{s(t-r)}}{s^{4/3}}\,ds .
\end{aligned}
\end{equation}
Defining $T=t-r$, the integral reduces to the standard inverse transform of $s^{-4/3}$ with a retarded shift.

The integrand
\begin{equation}
\label{LTintegrand}
\frac{e^{sT}}{s^{4/3}}
\end{equation}
has a branch point at $s=0$. Choosing the principal branch $-\pi<\arg s<\pi$, the branch cut lies along the negative real axis $(-\infty,0]$. For $T>0$, the Bromwich contour can be closed in the left half-plane, ensuring convergence due to the exponential damping of $e^{sT}$.

Let $C$ denote the closed contour consisting of the Bromwich line $C_B$, a large closing arc in the left half-plane, a small circle around the branch point, and the upper and lower sides of the branch cut.
\begin{figure}[t]
\centering
\includegraphics[width=0.9\columnwidth]{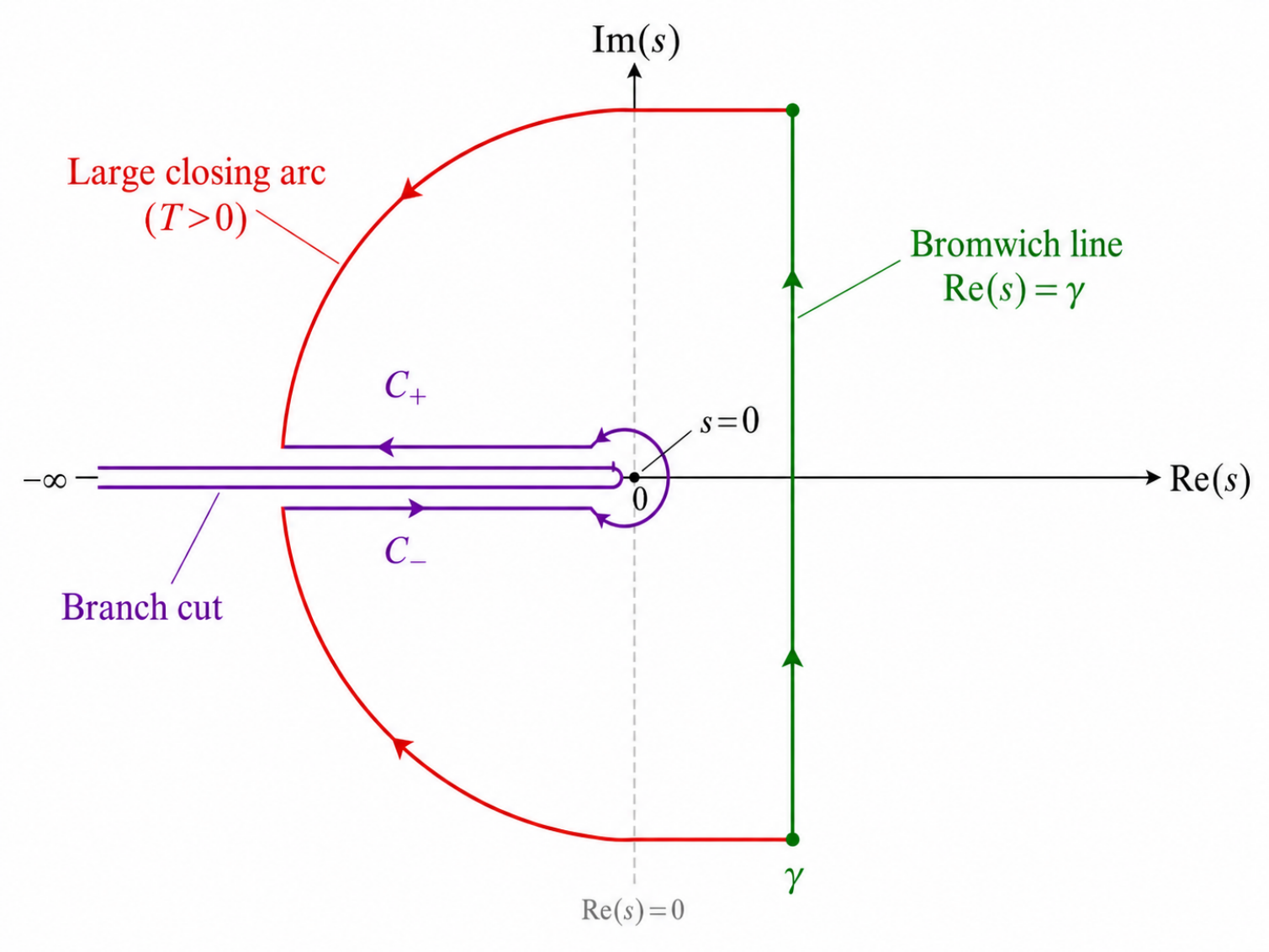}
\caption{Bromwich contour for the inverse Laplace transform of $s^{-4/3}$. The integrand has a branch point at $s=0$, with the branch cut chosen along the negative real axis.}
\label{fig:image5}
\vspace{-5pt}
\end{figure}
The contributions from the large closing arc and the small circle around $s=0$ vanish. Since there are no poles enclosed by the contour,
\begin{equation}
\label{LTclosedcontour}
\oint_C \frac{e^{sT}}{s^{4/3}}\,ds=0 .
\end{equation}
Thus,
\begin{equation}
\label{LTbromwichBranchRelation}
\int_{C_B}=-\left(\int_{C_+}+\int_{C_-}\right).
\end{equation}

On the upper side of the branch cut, let $s=x+i0^+$ with $x<0$, and write $x=-\rho$ with $\rho>0$. Then
\begin{equation}
\label{LTupperParam}
s=\rho e^{i\pi}, \qquad ds=-d\rho .
\end{equation}
Hence,
\begin{equation}
\label{LTupper}
\int_{C_+}=\int_{\infty}^{0}
e^{-\rho T}\rho^{-4/3}e^{-i4\pi/3}(-d\rho).
\end{equation}
Similarly, on the lower side,
\begin{equation}
\label{LTlowerParam}
s=\rho e^{-i\pi},
\end{equation}
giving
\begin{equation}
\label{LTlower}
\int_{C_-}=\int_{0}^{\infty}
e^{-\rho T}\rho^{-4/3}e^{i4\pi/3}d\rho .
\end{equation}
Combining the two contributions,
\begin{equation}
\label{LTdiscPre}
\int_{C_+}+\int_{C_-}
=\left(e^{i4\pi/3}-e^{-i4\pi/3}\right)
\int_{0}^{\infty}e^{-\rho T}\rho^{-4/3}d\rho .
\end{equation}
Using
\begin{equation}
\label{LTsineIdentity}
e^{ix}-e^{-ix}=2i\sin x,
\end{equation}
we obtain
\begin{equation}
\label{LTdisc}
\int_{C_+}+\int_{C_-}
=2i\sin\!\left(\frac{4\pi}{3}\right)
\int_{0}^{\infty}e^{-\rho T}\rho^{-4/3}d\rho .
\end{equation}
The remaining integral is evaluated using the Gamma-function identity
\begin{equation}
\label{LTgammaIdentity}
\int_0^\infty x^{\alpha-1}e^{-ax}dx=\frac{\Gamma(\alpha)}{a^\alpha}.
\end{equation}
With $\alpha=-1/3$, this gives
\begin{equation}
\label{LTgammaEval}
\int_0^\infty e^{-\rho T}\rho^{-4/3}d\rho=\Gamma(-1/3)T^{1/3}.
\end{equation}
Combining the branch-cut discontinuity with the Bromwich prefactor yields the standard result
\begin{equation}
\label{LTstandard}
\frac{1}{2\pi i}\int_{\gamma-i\infty}^{\gamma+i\infty}
\frac{e^{sT}}{s^{4/3}}\,ds
=\frac{T^{1/3}}{\Gamma(4/3)}\Theta(T).
\end{equation}
Substituting $T=t-r$ into \eqref{LTbromwich}, the time-domain radiative fields become
\begin{equation}
\label{invLapMagField}
B_{\hat{\theta}}(t)=E_{\hat{\phi}}(t)
=\frac{2B_0m_0}{r}
\left(-\frac{1}{\tau_{\rm H}}\right)^{1/3}
(t-r)^{1/3}\Theta(t-r)\sin\theta .
\end{equation}

The Heaviside factor $\Theta(t-r)$ explicitly encodes causality. An observer at radius $r$ detects no radiative signal for $t<r$, since the outgoing disturbance has not yet had time to propagate from the source to that radius. For $t>r$, the field turns on with the retarded-time dependence $(t-r)^{1/3}$. This is the Laplace-space analogue of imposing the outgoing-wave prescription in Fourier space. Without such a prescription, the field equations admit both incoming and outgoing branches. The one-sided Laplace transform, together with the Bromwich contour chosen for $T=t-r>0$, selects the causal outgoing branch appropriate to an evaporating black hole. This parallels the Landau prescription, where the physical solution is selected by the source turn-on condition and the outgoing radiation condition \cite{LandauLifshitzContinuousMedia}. Thus the arrow of emission is fixed by the retarded boundary condition on the solution, not by any modification of Maxwell's equations themselves.

Now, taking Fourier transformations of the magnetic and electric fields from \eqref{invLapMagField}, we get
\begin{equation}
\label{MF}
B_{\hat{\theta}}(\omega)=E_{\hat{\phi}}(\omega)
=
C_1
\frac{
B_0m_0e^{ir\omega}\sin\theta}
{r\,\tau_{\rm H}^{1/3}\omega^{4/3}},
\end{equation}
where $C_1=-2\Gamma\left(\frac{4}{3}\right)$ is a constant. Hence, for large $r$, the Poynting vector,
\begin{equation*}
\mathbf{S}(\omega)=\frac{1}{4\pi}\mathbf{E}(\omega)\times\mathbf{B}^{*}(\omega),
\end{equation*}
becomes
\begin{equation}
\label{(12)}
\mathbf{S}(\omega)=
\left\{
\frac{|C_1|^2}{4\pi}
\frac{
B_0^2m_0^2\sin^2\theta}
{r^2\tau_{\rm H}^{2/3}\omega^{8/3}},\,
\mathcal{O}\!\left(\frac{1}{r^3}\right),\,
0
\right\}.
\end{equation}
In the large $r$ limit, we drop any terms in the Poynting vector having $r$ dependence greater than $1/r^2$, since such terms decay rapidly for large $r$. Thus, $S_{\hat\theta}(\omega)\approx0$ and $S_{\hat r}(\omega)$ is the only non-zero component, as expected.

From the above Poynting vector, the emitted spectral energy over the full evaporation history becomes
\begin{equation}
\label{spectralenergyHawking}
\begin{aligned}
\frac{d\mathcal{E}_B}{d\omega}
&=
\frac{1}{2\pi}
\int d\Omega\,
\mathbf{S}(\omega)\cdot r^2\hat{\mathbf r}\\
&=
C_2\,
\frac{B_0^2m_0^2}
{\tau_{\rm H}^{2/3}\omega^{8/3}},
\end{aligned}
\end{equation}
where $C_2$ is another constant given by
\begin{equation*}
C_2=\frac{1}{3\pi}C_1^2.
\end{equation*}

The spectrum is infrared dominated. For the lifetime-integrated total energy, the lowest physically resolved frequency is set by the finite duration of the evaporation process, so that $\omega_{\min}\sim1/\tau_{\rm H}$. Taking $\omega_{\max}\sim c/R_{Sch}$, the total emitted energy is
\begin{equation}
\label{totalenergyHawking}
\begin{aligned}
\mathcal{E}_B
&=
C_2\int_{\omega_{\min}}^{\omega_{\max}}
\frac{B_0^2m_0^2}
{\tau_{\rm H}^{2/3}\omega^{8/3}}\,d\omega\\
&=
\frac{3C_2}{5}
\frac{B_0^2m_0^2}{\tau_{\rm H}^{2/3}}
\left(\omega_{\min}^{-5/3}-\omega_{\max}^{-5/3}\right).
\end{aligned}
\end{equation}
For $\omega_{\min}\sim1/\tau_{\rm H}$, we have $\omega_{\max}\gg\omega_{\min}$, so the contribution from the lower-frequency mode dominates. This gives
\begin{equation}
\label{totalenergysimplified}
\mathcal{E}_B\approx
\frac{3C_2}{5}B_0^2m_0^2\tau_{\rm H}.
\end{equation}
The lifetime of the black hole in CGS units is
\begin{equation}
\label{lifetime}
\tau_{\rm H}=
\frac{5120\,\pi\,G^2m_0^3}{c^4\hbar}.
\end{equation}
After restoring the constants $c$ and $G$, \eqref{totalenergysimplified} yields the total emitted energy in CGS units:
\begin{equation}
\label{totalenergyHawkingCGS}
\mathcal{E}_B\approx
C_3\,
\frac{B_0^2G^4m_0^5}{c^7\hbar},
\end{equation}
where $C_3=3072\,\pi C_2$ is another numerical constant.

We now consider the power averaged over a finite observation time $T_{\rm obs}$. A finite observation window imposes a frequency resolution of order $1/T_{\rm obs}$. Therefore, the appropriate infrared cutoff is
\begin{equation}
\label{omegaIR}
\omega_{\rm IR}
\sim
\max\left(\frac{1}{\tau_{\rm H}},\frac{1}{T_{\rm obs}}\right).
\end{equation}
The observation-time averaged spectral power is then
\begin{equation}
\label{instspectralpower}
\frac{dP_B}{d\omega}
=
\frac{1}{T_{\rm obs}}
\frac{d\mathcal{E}_B}{d\omega}
=
C_2\,
\frac{B_0^2m_0^2}
{T_{\rm obs}\tau_{\rm H}^{2/3}\omega^{8/3}}.
\end{equation}
The total power averaged over the observation time is obtained by integrating over the resolved frequency band:
\begin{equation}
\label{HawkingPowerattimeT}
\begin{aligned}
P_B(T_{\rm obs})
&=
\int_{\omega_{\rm IR}}^{\omega_{\max}}
\frac{dP_B}{d\omega}\,d\omega\\
&=
\frac{3C_2}{5}
\frac{B_0^2m_0^2}
{T_{\rm obs}\tau_{\rm H}^{2/3}}
\left[
\omega_{\rm IR}^{-5/3}
-\omega_{\max}^{-5/3}
\right].
\end{aligned}
\end{equation}

For an intermediate observation time $T_{\rm obs}<\tau_{\rm H}$, the infrared cutoff is set by the observation window, $\omega_{\rm IR}\sim1/T_{\rm obs}$. If, for a chosen $T_{\rm obs}$, $\omega_{\max}=c/R_{Sch}\gg1/T_{\rm obs}$, then \eqref{HawkingPowerattimeT} reduces to
\begin{equation}
\label{intermediatePower}
P_B(T_{\rm obs})
\approx
\frac{3C_2}{5}
B_0^2m_0^2
\left(\frac{T_{\rm obs}}{\tau_{\rm H}}\right)^{2/3}.
\end{equation}
Thus, an observation shorter than the evaporation time measures a smaller averaged power than the lifetime-averaged value, suppressed by the factor $(T_{\rm obs}/\tau_{\rm H})^{2/3}$.

For observation times comparable to the evaporation timescale, $T_{\rm obs}\sim\tau_{\rm H}$, we have $\omega_{\rm IR}\sim1/\tau_{\rm H}\equiv\omega_{\min}$. The lifetime-averaged total power is therefore
\begin{equation}
\label{(16)}
L_B=
\frac{3C_2}{5}
\frac{B_0^2m_0^2}{\tau_{\rm H}^{5/3}}
\left(\omega_{\min}^{-5/3}-\omega_{\max}^{-5/3}\right).
\end{equation}
Numerically, $3C_2/5\approx0.2$. Thus,
\begin{equation}
\label{FinalPower}
L_B=
0.2\,
\frac{B_0^2m_0^2}{\tau_{\rm H}^{5/3}}
\left(\omega_{\min}^{-5/3}-\omega_{\max}^{-5/3}\right).
\end{equation}
Since the Schwarzschild radius is given by $R_{Sch}=2m_0$, the expression \textbf{in CGS units} becomes
\begin{equation}
\label{(17)}
L_B=
0.05\,
\frac{B_0^2R_{Sch}^2c}{\tau_{\rm H}^{5/3}}
\Big(\omega_{\min}^{-5/3}-\omega_{\max}^{-5/3}\Big).
\end{equation}
In the above expression, we have restored the constants $G$ and $c$. Since $L_B \propto \omega^{-5/3}$, the emitted power is concentrated at lower frequencies and is therefore red-dominated.

Consequently, the lifetime-averaged emission is governed by modes with periods comparable to the full lifetime of the black hole, indicating that the emission is controlled primarily by the long-duration phase of the evolution. Since $\omega_{\min}\ll\omega_{\max}$, from \eqref{(17)}, we have
\begin{equation}
\label{appPower}
L_B\approx0.05\,B_0^2R_{Sch}^2c.
\end{equation}

\section{Discussion}

\subsection{Quantitative Analysis}

From the preceding calculations, it is evident that when a Schwarzschild black hole loses mass due to Hawking radiation, the interaction between the background magnetic field and the time-dependent spacetime gives rise to an effective electromagnetic source. This generates an additional channel of energy emission, distinct from the standard Hawking flux, and originates from the coupling between the evolving geometry and the external magnetic field.

It is important, however, to emphasize the perturbative nature of this result. In the Hawking calculation, the mass evolution $M(t)=m_0(1-t/\tau_{\rm H})^{1/3}$ was imposed as prescribed boundary data, corresponding to the standard Hawking evaporation law. The electromagnetic radiation derived above is therefore a leading-order correction to that prescribed evolution. It cannot be allowed to dominate over the Hawking channel within the present treatment. If the electromagnetic energy loss became comparable to or larger than the Hawking energy loss, the mass-loss law itself would be modified, schematically as
\begin{equation}
\dot{M}=\dot{M}_{\rm H}+\dot{M}_{B},
\end{equation}
where $\dot{M}_{B}$ is the mass-loss correction due to the new electromagnetic emission channel. In that case, the assumed Hawking form of $M(t)$ would have to be solved for self-consistently. Thus, the regime of validity of the present calculation requires the induced electromagnetic channel to remain perturbative.

If the black hole were to decay purely via Hawking radiation, the entire rest mass energy would be lost into radiation, yielding
\begin{equation}
\mathcal{E}_{\rm H}=m_0c^2 .
\end{equation}
Comparing the total energy lost in the induced electromagnetic channel to the Hawking sector, we obtain from \eqref{totalenergyHawkingCGS}:
\begin{equation}
 \label{(44)b)}
 \begin{split}
 \frac{\mathcal{E}_B}{\mathcal{E}_{\rm H}}
 &\approx 3266\,  \frac{B_0^2\,G^4\,m_0^4}{c^9\,\hbar}\\
 &\approx3 \times10^{-93}B_0^2\,m_0^4.
\end{split}
\end{equation}
Here $B_0$ is measured in gauss and $m_0$ in grams in the second line. To better understand the scaling, it is useful to express this ratio in terms of fundamental quantities. Defining the Planck mass and a natural magnetic field scale as
\begin{equation}
M_P=\sqrt{\frac{\hbar c}{G}},
\qquad
B_P=\frac{c^{7/2}}{G\sqrt{\hbar}},
\end{equation}
equation \eqref{(44)b)} can be rewritten as
\begin{equation}
\frac{\mathcal{E}_B}{\mathcal{E}_{\rm H}}
\approx 1.3\times10^4\,
\left(\frac{B_0}{B_P}\right)^2
\left(\frac{m_0}{M_P}\right)^4.
\end{equation}

The perturbative consistency condition is therefore
\begin{equation}
\label{perturbativecondition}
\frac{\mathcal{E}_B}{\mathcal{E}_{\rm H}}\ll 1 .
\end{equation}
Equivalently, the boundary at which the electromagnetic channel would become comparable to the Hawking channel is
\begin{equation}
\label{power}
\begin{split}
m_0 & \sim  10^{23}\,B_0^{-1/2} \,\text{gram}\\
& \sim 10^{-10} M_\odot\,B_0^{-1/2}.
\end{split}
\end{equation}
This should be interpreted as an upper limit on the regime of validity of the present perturbative model, not as a regime in which the electromagnetic channel is allowed to dominate. For a weak background magnetic field of order $B_0\sim1$ gauss, self-consistency requires
\begin{equation}
m_0 \ll 10^{23}\,\text{gram}
\end{equation}
up to the order-unity numerical factors displayed above.

This perturbative energy bound is in addition to the weak-field assumptions made in constructing the metric and electromagnetic response. The background magnetic field must be sufficiently weak that its stress-energy does not backreact on the spacetime geometry. A rough estimate of this condition is that the magnetic energy contained in a region of size $R_{Sch}$ should be much smaller than the black hole rest energy,
\begin{equation}
\frac{B_0^2 R_{Sch}^3}{m_0c^2}\ll1,
\end{equation}
up to order-unity geometric factors. Equivalently,
\begin{equation}
B_0 \ll \frac{c^4}{G^{3/2}m_0}.
\end{equation}
For gauss-scale magnetic fields and masses satisfying \eqref{perturbativecondition}, this magnetic-backreaction constraint is easily satisfied. The stronger restriction in the present Hawking application comes from requiring the induced electromagnetic energy loss to remain a small correction to the assumed Hawking evolution.

The same point can be expressed through the total energy-loss law. Including the induced electromagnetic channel, the total energy loss is
\begin{equation}
\mathcal{E}_{\text{Total}}=-(\mathcal{E}_{\rm H}+\mathcal{E}_B)
\approx -m_0c^2\left[1+
3266\,\frac{B_0^2 G^4m_0^4}{c^9\,\hbar }\right].
\end{equation}
Within the regime of validity of the calculation, the term proportional to $B_0^2m_0^4$ must be much smaller than unity. Hence, the additional electromagnetic channel appears as a small correction to the standard Hawking evaporation law, governed by the low-frequency sector of the spectrum and by the slowly varying large-scale spacetime geometry.

Thus, although a magnetized environment induces an additional classical electromagnetic energy-loss channel, the present calculation is self-consistent only while that channel remains perturbative. If the induced electromagnetic luminosity became comparable to the Hawking luminosity, the central mass evolution would have to be recomputed with both channels included, and the prescribed Hawking mass profile used in the present work would no longer be valid. A fully self-consistent treatment of that non-perturbative regime is beyond the scope of this analysis.

\subsection{Transmission-line interpretation of the Hawking-driven electromagnetic radiation}
\label{sec:tl_analogy}

It is useful to reinterpret the Hawking-driven electromagnetic perturbation as a \emph{driven, lossless transmission line} with position-dependent reactive loading. In the black-hole setting, the interaction between the time-dependent spacetime and the background magnetic field acts as an effective distributed drive for the electromagnetic perturbation, generating an outgoing radiative electromagnetic mode. This mirrors a transmission line, where an external source feeds a distributed medium that stores energy locally and carries it as propagating waves to a load. In the far zone, the radiative Poynting flux corresponds to the power carried by traveling voltage and current waves and delivered to a matched load.

After angular decomposition, Maxwell's equations in \eqref{(7)} reduce to a one-dimensional radial equation with a source term and an effective potential barrier. In this picture, the Hawking-driven effective source acts as a distributed drive, the angular-momentum barrier appears as a spatially varying reactive element, and the outgoing radiation corresponds to the voltage delivered at the end of the line.

\begin{figure}[t]
\centering
\includegraphics[width=\columnwidth]{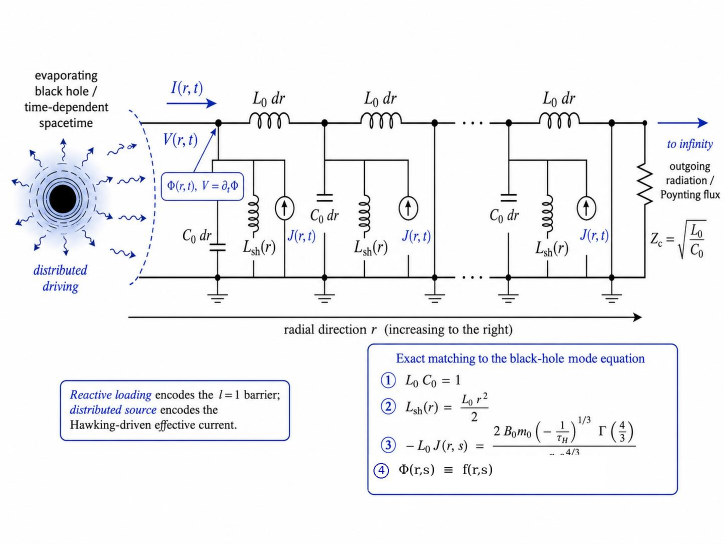}
\caption{Circuit analogue of the Hawking-driven electromagnetic radiation. The interaction between the evolving spacetime and the background magnetic field is represented by a distributed current source $J(r,t)$ feeding a lossless transmission line. The additional shunt inductance density $L_{\rm sh}(r)$ encodes the reactive barrier, and the outgoing radiation corresponds to the voltage delivered to a matched load at infinity.}
\label{fig:circuit_analogue}
\vspace{-5pt}
\end{figure}

We begin from the standard lossless transmission-line model. Consider a line parameterized by the radial coordinate $r$, with voltage $V(r,t)$ and current $I(r,t)$. An infinitesimal segment $dr$ consists of a series inductance $L_0\,dr$ and a shunt capacitance $C_0\,dr$ to ground. The telegrapher equations are
\begin{equation}
\partial_r V(r,t)=-L_0\,\partial_t I(r,t),
\label{TL1}
\end{equation}
\begin{equation}
\partial_r I(r,t)=-C_0\,\partial_t V(r,t).
\label{TL2}
\end{equation}
Equation \eqref{TL1} represents the inductive response to a time-varying current, while \eqref{TL2} is Kirchhoff's current law for the shunt capacitor. Together they describe free wave propagation through exchange of magnetic and electric energy along the line.

The Hawking-driven radial equation is not a free-line equation. Compared with a free one-dimensional wave equation, \eqref{(7)} contains a position-dependent barrier term $2f/r^2$ and a distributed source term proportional to $1/(r\,s^{4/3})$. The circuit analogue therefore modifies the line by adding an additional \emph{shunt inductance density} $L_{\rm sh}(r)$ and an external distributed current source $J(r,t)$ at each point. The shunt inductance density stores magnetic energy locally and modifies propagation without dissipation, while the distributed source represents the effective current generated by the coupling between the time-dependent geometry and the background magnetic field.

Introducing the node-flux variable
\begin{equation}
\Phi(r,t)\equiv \int^t V(r,t')\,dt',
\qquad \Rightarrow \qquad
V(r,t)=\partial_t \Phi(r,t),
\label{fluxdef}
\end{equation}
allows the dynamics to be written in terms of a single field. From \eqref{TL1}, one obtains
\begin{equation}
I(r,t)=-\frac{1}{L_0}\,\partial_r \Phi(r,t),
\label{currentphi}
\end{equation}
so voltage and current correspond to temporal and spatial variations of $\Phi$.

Applying Kirchhoff's current law at each node, the currents to ground consist of the capacitor current $C_0\partial_t V$, the shunt-inductor current, and the externally injected current. For the shunt inductive loading,
\begin{equation}
V(r,t)=L_{\rm sh}(r)\,\partial_t I_{\rm sh}(r,t),
\end{equation}
which, using \eqref{fluxdef}, integrates to
\begin{equation}
I_{\rm sh}(r,t)=\frac{\Phi(r,t)}{L_{\rm sh}(r)}.
\label{Ish}
\end{equation}
Hence the modified telegrapher equation is
\begin{equation}
\partial_r I(r,t)
=
-\,C_0\,\partial_t V(r,t)
-\frac{\Phi(r,t)}{L_{\rm sh}(r)}
+J(r,t),
\label{modTL2}
\end{equation}
where the second term is the shunt-inductor current density and the third is the external distributed source.

Substituting \eqref{fluxdef} and \eqref{currentphi} into \eqref{modTL2} gives
\begin{equation}
\partial_r^2\Phi
-
L_0C_0\,\partial_t^2\Phi
-
\frac{L_0}{L_{\rm sh}(r)}\,\Phi
=
-\,L_0J(r,t).
\label{TLwave_t}
\end{equation}
The term proportional to $1/L_{\rm sh}(r)$ acts as a position-dependent reactive barrier, while $J(r,t)$ is a distributed source.

After Laplace transform in time, \eqref{TLwave_t} becomes
\begin{equation}
\Phi''(r,s)-L_0C_0\,s^2\Phi(r,s)-\frac{L_0}{L_{\rm sh}(r)}\Phi(r,s)
=
-\,L_0J(r,s).
\label{TLwave_s}
\end{equation}
Choosing
\begin{equation}
L_0C_0=1,
\label{match1}
\end{equation}
\begin{equation}
\frac{L_0}{L_{\rm sh}(r)}=\frac{2}{r^2}
\qquad\Longrightarrow\qquad
L_{\rm sh}(r)=\frac{L_0r^2}{2},
\label{match2}
\end{equation}
and
\begin{equation}
-L_0J(r,s)
=
\frac{2B_0m_0\left(-\frac{1}{\tau_{\rm H}}\right)^{1/3}\Gamma(4/3)}{r\,s^{4/3}},
\label{match3}
\end{equation}
reduces \eqref{TLwave_s} exactly to \eqref{(7)}, with
\begin{equation}
\Phi(r,s) \equiv f(r,s).
\label{phiequalsf}
\end{equation}
Thus the black-hole mode equation is reproduced by a driven transmission line. The source dependence $J(r,s)\propto s^{-4/3}/r$ corresponds to a fractional-integrator response applied to a broadband drive: an ordinary integrator ($1/s$) cascaded with a fractional integrator of order $1/3$, approximated over a finite bandwidth by an RC ladder with logarithmically spaced time constants. The factor $1/r$ is implemented by distributing the injected waveform to each node with amplitude proportional to $1/r$.

The matched elements have a clear physical meaning. The shunt inductance density $L_{\rm sh}(r)$ is small near the source and large at infinity, reproducing the $\ell=1$ barrier, while $J(r,t)$ continuously injects energy along the line with spatial weighting.

Finally, the far-zone observable is the outgoing voltage on the asymptotically free line. For large $r$, the outgoing solution from \eqref{(7)} in Fourier domain behaves as
\begin{equation}
\Phi_{\rm out}(r,\omega)\propto \omega^{-7/3}\,e^{i\omega r},
\end{equation}
so from \eqref{fluxdef}
\begin{equation}
V_{\rm out}\propto \omega\,\Phi_{\rm out}
\quad\Rightarrow\quad
V_{\rm out}\propto \omega^{-4/3}\,e^{i\omega r}.
\end{equation}
For a matched load with $Z_c=\sqrt{L_0/C_0}$, the spectral power density is
\begin{equation}
\frac{dP_{\rm TL}}{d\omega}\propto \frac{|V_{\rm out}(\omega)|^2}{Z_c}\propto \omega^{-8/3}.
\label{PTL}
\end{equation}
Thus the transmission-line analogue reproduces the same scaling as the Hawking spectrum,
\begin{equation}
\frac{d P_B}{d\omega}\propto \omega^{-8/3}.
\end{equation}

In this way, the evaporating black hole acts as a distributed source feeding a lossless transmission line, with the barrier encoded in reactive loading and the observed radiation corresponding to the delivered voltage at infinity.

\subsection{Electromagnetic Emission from an Evaporating Black Hole: A Macroscopic Medium Perspective}

In general relativity, electromagnetic wave propagation in a gravitational field can be understood through an optical-mechanical analogy. In the classic treatment of Landau and Lifshitz, a gravitational field may be represented as an effective optical medium, so that light propagation in curved spacetime resembles propagation through an inhomogeneous material with geometry-dependent dielectric properties \cite{LandauLifshitz}. This picture is especially intuitive for stationary geometries, where the gravitational field acts as a fixed optical medium.

Here we use the related but more directly dynamical $3+1$ formulation. The ADM decomposition foliates spacetime into spatial hypersurfaces and separates the lapse, shift, and spatial metric variables \cite{ADM1962}. In this language, Komissarov's formulation rewrites Maxwell's equations in terms of three-dimensional fields measured on the spatial slices, with the lapse, shift, and spatial metric entering the effective constitutive relations of the vacuum \cite{Komissarov}. Equivalently, the curved vacuum behaves as an effective optical medium whose permittivity and permeability are determined by the geometry. This is the appropriate framework for the present problem, because the lapse $\alpha(r,t)$ is not merely spatially inhomogeneous but explicitly time-dependent due to evaporation.

To construct this picture, we use the $3+1$ decomposition
\[
ds^2 = -\alpha^2 dt^2 + \gamma_{ij} (dx^i + \beta^i dt) (dx^j + \beta^j dt),
\]
where $\gamma_{ij}$ is the spatial metric, $\alpha$ is the lapse function, and $\beta^i$ is the shift vector. For a non-rotating Schwarzschild black hole, $\beta^i = 0$. The lapse function follows from the weak-field metric component $g_{tt} \approx -1 + 2m(r,t)/r - 2m(r,t)/r_0$:
\begin{equation}
\label{K1}
    \alpha = \sqrt{-g_{tt}} \approx 1 - \frac{m(r,t)}{r} + \frac{m(r,t)}{r_0}
\end{equation}

Using the vacuum constitutive equations derived by Komissarov \cite{Komissarov},
\begin{equation}
    \mathbf{E} = \alpha \mathbf{D} + \bm{\beta} \times \mathbf{B}, \quad \mathbf{H} = \alpha \mathbf{B} - \bm{\beta} \times \mathbf{D}
\end{equation}
and setting $\beta^i = 0$, we obtain
\begin{equation}
   \mathbf{D} = \alpha^{-1} \mathbf{E} , \quad \mathbf{B} = \alpha^{-1} \mathbf{H}
\end{equation}
Thus the vacuum behaves as an isotropic effective medium with scalar permittivity and permeability:
\begin{equation}
    \epsilon(r,t) = \mu(r,t) = \frac{1}{\alpha(r,t)} \approx 1 + \frac{m(r,t)}{r} - \frac{m(r,t)}{r_0}.
\end{equation}
The distinction from the usual stationary optical-medium analogy is important: here $\epsilon$ and $\mu$ depend on time through $m(r,t)$. Therefore the medium does not merely bend or delay electromagnetic fields; it relaxes as the black hole evaporates. The interaction between this time-dependent effective medium and the pre-existing background magnetic field generates an effective macroscopic current, which acts as the source of the outgoing electromagnetic perturbation.

\begin{figure}[t]
\centering
\includegraphics[width=\columnwidth]{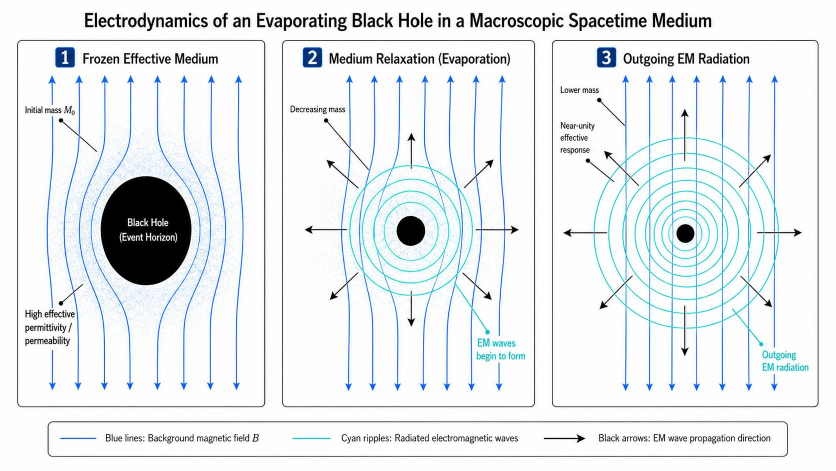}
\caption{Macroscopic-medium interpretation of the emission process. Initially, the black hole produces a strong effective permittivity/permeability response, distorting the background magnetic field. During evaporation, the decreasing mass relaxes this effective medium toward the flat-space value. The interaction between the relaxing medium and the background magnetic field generates an effective displacement-current-like source, producing outgoing electromagnetic waves. Blue curves denote the background magnetic field, cyan ripples denote the radiated electromagnetic waves, and white arrows denote the EM-wave propagation direction.}
\label{fig:macroscopic_medium}
\vspace{-5pt}
\end{figure}

Figure~\ref{fig:macroscopic_medium} summarizes this physical picture. In the initial stage, the larger black-hole mass produces stronger effective permittivity and permeability and distorts the background magnetic field. As evaporation proceeds, the effective medium relaxes toward the flat-space value while the background field readjusts. The resulting time-dependent interaction produces outgoing electromagnetic radiation.

The time dependence of the lapse modifies the electromagnetic problem through Maxwell's equations on the evolving spatial slices. The exact covariant equations give the $3+1$ Ampere law
\begin{equation}
    \frac{1}{\sqrt{\gamma}} \partial_t (\sqrt{\gamma} D^i) - (\nabla \times H)^i = 0.
\end{equation}
For a stationary effective medium this equation would describe field propagation in a fixed optical background. In the present case, however, $\alpha(r,t)$ and hence the constitutive response are time-dependent. Decomposing the fields into a static background and a wave perturbation, $D = \delta D$ and $H = H^{(0)} + \delta H$, isolates wave propagation on the left and the effective macroscopic source on the right:
\begin{equation}
\label{Komi}
    \frac{1}{\sqrt{\gamma}} \partial_t (\sqrt{\gamma} \delta D^i) - (\nabla \times \delta H)^i = (\nabla \times H^{(0)})^i \equiv J_{\text{eff}}^i .
\end{equation}

The uniform external magnetic field is described by
\[
A_\phi^{(0)} = \frac{1}{2}B_0 r^2 \sin^2\theta .
\]
On the unperturbed spatial metric, $\gamma_{rr}=1$, $\gamma_{\theta\theta}=r^2$, $\gamma_{\phi\phi}=r^2\sin^2\theta$, with $\sqrt{\gamma}=r^2\sin\theta$, the contravariant magnetic field is
\[
B^i = \frac{1}{\sqrt{\gamma}} \epsilon^{ijk} \partial_j A_k^{(0)}.
\]
Evaluating the non-zero components gives $B^r = B_0 \cos\theta$ and $B^\theta = -B_0 \sin\theta / r$. Lowering the indices gives $B_r = B_0 \cos\theta$ and $B_\theta = -B_0 r \sin\theta$.

The macroscopic background fields follow from $H_k^{(0)} = \alpha B_k$:
\begin{align}
    H_r^{(0)} &= \alpha B_0 \cos\theta \\
    H_\theta^{(0)} &= -\alpha B_0 r \sin\theta
\end{align}
The source current requires the curl of this background field. Defining $C_\phi = \partial_r H_\theta^{(0)} - \partial_\theta H_r^{(0)}$, we have
\begin{align}
    \partial_\theta H_r^{(0)} &= -\alpha B_0 \sin\theta \\
    \partial_r H_\theta^{(0)} &= -B_0 \sin\theta \partial_r(r\alpha)
\end{align}
and therefore
\begin{equation}
    C_\phi = -B_0 \sin\theta \left[ \partial_r (r\alpha) - \alpha \right] = -B_0 \sin\theta (r \partial_r \alpha)
\end{equation}
Since $\partial_r \alpha \sim \mathcal{O}(m)$, this curl is entirely first order in the metric perturbation.

The true contravariant current is obtained by dividing by the perturbed spatial volume element $\sqrt{\gamma} \approx r^2\sin\theta\left(1 + \frac{3m}{r}\right)$:
\begin{equation}
    J_{\text{eff}}^\phi = \frac{-B_0 \sin\theta (r \partial_r \alpha)}{r^2\sin\theta\left(1 + \frac{3m}{r}\right)} \approx - \frac{B_0 \partial_r \alpha}{r} \left(1 - \frac{3m}{r}\right)
\end{equation}
Because the numerator is already $\mathcal{O}(m)$, the volume correction contributes only at $\mathcal{O}(m^2)$ and is discarded at leading order. Thus,
\begin{equation}
    J_{\text{eff}}^\phi = - \frac{B_0 \partial_r \alpha}{r}
\end{equation}
with $\alpha$ given by \eqref{K1}. Evaluating this source in the near-zone ($r \ll r_0$) and transforming to Laplace space, using $m(r,s)$ from \eqref{21}, gives the scaling
\begin{equation}
    J_{\text{eff}}^\phi(s) \propto
    \frac{B_0 m_0}{r^3}
    \frac{
    \left(-\frac{1}{\tau_{\rm H}}\right)^{1/3}
    \Gamma\left(\frac{4}{3}\right)}
    {s^{4/3}} .
\end{equation}
The simplified macroscopic-medium picture captures the physical origin of the source and its low-frequency scaling, while the exact numerical normalization is fixed by the full covariant source term derived in Sec.~\ref{source}.

This construction separates the perturbative scales clearly. The background magnetic field is order $\mathcal{O}(1)$, while the relaxing geometry introduces an $\mathcal{O}(m)$ perturbation. Their interaction generates $J_{\text{eff}}^\phi$, and hence the wave response $(\delta A_\phi, \delta D, \delta H)$, at order $\mathcal{O}(m)$. Any coupling between these wave fields and the $\mathcal{O}(m)$ corrections in $\gamma_{ij}$ or $\alpha$ would be $\mathcal{O}(m^2)$ and is consistently neglected. Therefore, the wave propagation on the left-hand side of \eqref{Komi} is evaluated on the unperturbed flat background, $\gamma_{ij}\to\eta_{ij}$, $\alpha=1$, and $\epsilon=1$.

Using the transverse ansatz $\delta A_\phi(r,t)=f(r,t)\sin^2\theta$, the displacement current becomes $\delta D^\phi=\delta E^\phi=-\partial_t f/r^2$, so
\begin{equation}
    \frac{1}{\sqrt{\gamma}} \partial_t (\sqrt{\gamma} \delta D^\phi) = \frac{1}{r^2\sin\theta} \partial_t \left( r^2\sin\theta \left[ -\frac{\partial_t f}{r^2} \right] \right) = -\frac{\partial_t^2 f}{r^2}
\end{equation}
The magnetic perturbation $\delta B^i = \frac{1}{\sqrt{\gamma}} \epsilon^{ijk} \partial_j \delta A_k$ has non-zero components
\begin{equation}
    \delta B^r = \frac{2f\cos\theta}{r^2}, \quad \delta B^\theta = -\frac{\partial_r f \sin\theta}{r^2}
\end{equation}
and, since $\alpha=1$ and $\mu=1$ for the propagating perturbation,
\begin{equation}
    \delta H_r = \frac{2f\cos\theta}{r^2}, \quad \delta H_\theta = -\partial_r f \sin\theta
\end{equation}
The macroscopic magnetic curl is then
\begin{equation}
\begin{aligned}
    (\nabla \times \delta H)^\phi
    &= \frac{1}{r^2\sin\theta} \left[ -\partial_r^2 f \sin\theta
       + \frac{2f\sin\theta}{r^2} \right] \\
    &= -\frac{\partial_r^2 f}{r^2} + \frac{2f}{r^4}.
\end{aligned}
\end{equation}
Combining this with the temporal operator gives the complete $\ell=1$ wave operator:
\begin{equation}
    \frac{1}{\sqrt{\gamma}} \partial_t (\sqrt{\gamma} \delta D^\phi)-(\nabla \times \delta H)^\phi = \frac{1}{r^2} \left[ \partial_r^2 f - \partial_t^2 f - \frac{2}{r^2} f \right]
\end{equation}

Equating this to $J_{\text{eff}}^\phi(s)$ and multiplying by $r^2$ gives a radial inhomogeneous equation with the same radial and frequency dependence as the covariant calculation:
\begin{equation}
    \partial_r^2 f(r,s)
    - \left( \frac{2}{r^2} + s^2 \right) f(r,s)
    \propto{} \frac{B_0 m_0}{r s^{4/3}} 
   \left(-\frac{1}{\tau_{\rm H}}\right)^{1/3}
    \Gamma\!\left(\frac{4}{3}\right)
\end{equation}
Solving with outgoing boundary conditions gives $\delta A_\phi(s) \propto e^{-rs}/(r s^{7/3})$, and the far-zone transverse fields are
\begin{equation}
    B_\theta(s) = E_\phi(s) \propto \frac{e^{-rs}}{r s^{4/3}}
\end{equation}
The inverse Laplace transform gives the retarded-time profile $B_\theta(t) \propto (t-r)^{1/3} \Theta(t-r)$, and the Fourier transform gives
\begin{equation}
    B_\theta(\omega) = E_\phi(\omega) \propto \omega^{-4/3}
\end{equation}
The radial Poynting flux is $S_r(\omega)=\frac{1}{4\pi}|E_\phi(\omega)|^2$, so
\begin{equation}
   S_r(\omega) \propto  \omega^{-8/3}
\end{equation}
and therefore
\begin{equation}
   \frac{d\mathcal{E}_B}{d\omega} \propto \omega^{-8/3}
\end{equation}

Thus, the macroscopic-medium picture reproduces the same $\omega^{-8/3}$ spectrum obtained from the covariant curved-spacetime calculation. In this interpretation, the evaporating geometry acts as a time-dependent effective dielectric; its relaxation in the presence of the background magnetic field produces an effective macroscopic current that drives outgoing electromagnetic radiation.

\section{Conclusion}

We have shown that an evaporating, non-rotating black hole placed in an external magnetic field develops an effective electromagnetic source through the time dependence of the spacetime geometry. Starting from a weak-field, spherically symmetric metric with prescribed Hawking mass loss, we derived the corresponding source term in Maxwell's equations and solved for the induced far-zone radiation. The resulting emission is a classical electromagnetic correction to the standard Hawking evaporation process, rather than Hawking radiation itself.

The main result is that the Hawking-driven source produces a broadband, infrared-dominated spectrum. In the far zone, the leading radiative fields scale as
\begin{equation}
E_\phi(\omega)=B_\theta(\omega)\propto \omega^{-4/3},
\end{equation}
so that
\begin{equation}
\frac{d\mathcal{E}_B}{d\omega}\propto \omega^{-8/3}.
\end{equation}
The emission is therefore controlled by long-duration modes associated with the slow evaporation timescale. The inverse Laplace construction selects the retarded outgoing branch, giving time-domain fields proportional to $\Theta(t-r)$ and fixing the causal arrow of emission.

The polarization structure is fixed by the geometry of the perturbation. For radial propagation, the leading outgoing wave is transverse, with electric field along the azimuthal direction and magnetic field along the polar direction. Since $E_r=0$, the radiation is a Transverse Electric (TE) mode with respect to the radial propagation direction. Equivalently, the axial vector-potential perturbation produces the radiative pair $E_\phi$ and $B_\theta$, orthogonal both to each other and to the direction of propagation.

The auxiliary examples confirm that the same construction applies beyond the Hawking profile. The oscillating case gives monochromatic emission, while the exponentially decaying case provides a Laplace-space check of the outgoing-boundary prescription. The Vaidya and Oppenheimer--Snyder comparisons show that time dependence alone is not sufficient: a nontrivial radial structure in the effective mass profile is required to generate the electromagnetic source.

The induced electromagnetic channel must be interpreted perturbatively. In the Hawking case, the mass profile
\begin{equation}
M(t)=m_0\left(1-\frac{t}{\tau_{\rm H}}\right)^{1/3}
\end{equation}
was imposed as prescribed boundary data. Therefore the electromagnetic energy loss cannot dominate over the Hawking channel within the present treatment. If the induced luminosity became comparable to the Hawking luminosity, the mass evolution would have to be recomputed self-consistently with both channels included. The bounds derived in the discussion therefore define the validity regime of the perturbative model, not a regime in which the new channel replaces Hawking evaporation.

This mechanism is conceptually reminiscent of the Gertsenshtein--Zel'dovich effect, in the sense that a background magnetic field mediates the conversion of gravitational dynamics into electromagnetic radiation \cite{gertsenshtein1,gertsenshtein2}. However, the present mechanism is distinct from the standard Gertsenshtein--Zel'dovich conversion process. Unlike the (time-reversible) EM-GW scattering process, we explore an intrinsic macroscopic emission process. Here the source is not an incident gravitational wave being converted into an electromagnetic wave; it is the radial, time-dependent mass profile of the evaporating geometry. The evolving spacetime makes the otherwise static magnetic background fail to satisfy Maxwell's equations independently, thereby inducing an effective electromagnetic source proportional to the radial structure of $m(r,t)$.

The macroscopic-medium and transmission-line interpretations give complementary pictures of the same result. In the medium picture, the evaporating geometry acts as a time-dependent effective dielectric whose relaxation, in the presence of the background magnetic field, generates an effective macroscopic current. In the transmission-line picture, the same dynamics is represented by a distributed drive feeding a lossless line with position-dependent reactive loading. Both viewpoints reproduce the infrared scaling and clarify how the evolving geometry stores, releases, and transports electromagnetic energy.

Several extensions remain natural. A fully self-consistent treatment would include the induced electromagnetic channel directly in the mass-loss equation, allowing the central mass to evolve under both the Hawking flux and the additional electromagnetic luminosity. It would also be important to go beyond the weak-field approximation, include magnetic backreaction, and generalize the calculation to rotating black holes, where frame dragging, spin-induced multipoles, and magnetospheric structure may modify both the source term and the polarization pattern.

Possible astrophysical settings include primordial black holes in ambient interstellar or primordial magnetic fields, compact objects undergoing slow spherical accretion or mass loss, nearly spherical collapse or rebound scenarios, and black holes embedded in magnetized plasma environments such as accretion flows, neutron-star neighborhoods, or early-universe magnetized regions. In each case, the effect is controlled by the rate of mass evolution, the radial structure of the outgoing energy flux, and the strength and geometry of the background magnetic field. Determining whether the resulting low-frequency, TE-polarized electromagnetic component can leave an observable imprint remains an important question for future work.

\section*{Acknowledgment}
The authors would like to thank Eric Poisson and Sergei Khlebnikov for their valuable comments and suggestions on improving the manuscript.

\appendix

\section*{Appendix organization}
The appendices collect auxiliary examples that support the calculation for Hawking radiation in the main text. Appendix~\ref{oscillate} treats the oscillating-mass case, with the Green's function derivation as the main calculation and the small-$r$, small-$\Omega$ source expansion as a complementary check. Appendix~\ref{Laplace} treats the exponentially decaying trial case. Appendices~\ref{Vaidya} and~\ref{Oppenheimer} discuss why the Vaidya and Oppenheimer--Snyder geometries do not generate a non-vanishing electromagnetic source in the present construction.

\section{Trial Case I: Black hole with oscillating mass}
\label{oscillate} 

In this appendix, we consider a simplified scenario where the central mass undergoes small, slow oscillations:
\begin{equation}
\label{osc}
m(r=2m_0,t) = m_0[1 + \epsilon\cos(\Omega t)]
\end{equation}
with amplitude $\epsilon \ll 1$ and frequency $\Omega \ll 1/m_0$. We use the asymmetric Fourier convention
\begin{equation}
\label{o1}
\begin{aligned}
m(r,\omega)&=\int_{-\infty}^{\infty} e^{i \omega t} m(r,t)\,dt,\\
m(r,t)&=\frac{1}{2\pi}\int_{-\infty}^{\infty} e^{-i \omega t} m(r,\omega)\,d\omega .
\end{aligned}
\end{equation}
Equation \eqref{10} becomes
\begin{equation}
\label{o1b}
r\,m''(r,\omega)+m'(r,\omega)+2r_0\,\omega^2\,m(r,\omega)=0 .
\end{equation}
The general solution is
\begin{equation}
\label{o3}  
m(r,\omega)=c_1(\omega)\,H_0^{(1)}\left(2\omega \sqrt{2 r r_0}\right)+ c_2(\omega)\, H_0^{(2)}\left(2\omega \sqrt{2 r r_0}\right),
\end{equation}
where $H_0^{(1)}$ and $H_0^{(2)}$ are Hankel functions of the first and second kinds. Since the central mass is oscillatory, the auxiliary source alternates between phases of decreasing and increasing mass. In the decreasing phase, the source may be interpreted as emitting an outgoing null perturbation carrying positive energy density. In the increasing phase, within the same linearized trial construction, the corresponding perturbation carries negative energy density. In both cases, the physical prescription is that the disturbance is generated at the central object and propagates outward, rather than being incident from spatial infinity. We therefore impose an outgoing-wave boundary condition at infinity, which removes the incoming Hankel branch and requires $c_2(\omega)=0$. The solution then reduces to
\begin{equation}
\label{o4}
m(r,\omega)=c_1(\omega)\,H_{0}^{(1)}\left(2\omega\sqrt{2rr_{0}}\right).
\end{equation}

Taking the Fourier transform of Eq.~\eqref{osc} for non-zero $\omega$ gives
\begin{equation}
\label{o5}
m(r=2m_0,\omega)=m_0\pi\epsilon \big[ \delta(\omega - \Omega)+\delta(\omega + \Omega)\big].
\end{equation}
The delta functions in \eqref{o5} restrict the solution to $\omega=\pm\Omega$. For compactness, the notation $\left. X(\omega)\right|_{\omega=\Omega}$ denotes $X(\Omega)\delta(\omega-\Omega)$, while $\left. X(\omega)\right|_{\omega=-\Omega}$ denotes $X(-\Omega)\delta(\omega+\Omega)$. With this convention, matching Eq.~\eqref{o4} to \eqref{o5} at $r=2m_0$ determines
\begin{equation}
\label{o6}
c_1(\omega)=\left.\frac{ m_{0}\pi\epsilon} { H_{0}^{(1)}\left( 4\omega \sqrt{m_{0}r_0} \right) }\right|_{\omega =\pm \Omega} .
\end{equation}
Therefore, 
\begin{equation}
\label{o7}
\begin{split}
m(r,\omega) &= \left.\frac{m_0\pi\epsilon\,H_{0}^{(1)}\left( 2\Omega \sqrt{2rr_0} \right) }{H_{0}^{(1)}\left( 4\Omega \sqrt{m_{0}r_0} \right)}\right|_{\omega = \Omega} \\
&\quad + \left.\frac{m_0\pi\epsilon\,H_{0}^{(1)}\left( -2\Omega \sqrt{2rr_0} \right) }{H_{0}^{(1)}\left( -4\Omega \sqrt{m_{0}r_0} \right)}\right|_{\omega =- \Omega}.
\end{split}
\end{equation}

The characteristic scale $r_0$ is fixed from the slow-oscillation limit. The leading term in the small-$\Omega$ expansion is independent of $r_0$, so the first nontrivial radial term gives
\begin{equation}
\label{o10}
\frac{ m_{0}\pi\epsilon\, H_{0}^{(1)}\left(\pm 2\Omega \sqrt{2rr_0} \right)}{ H_{0}^{(1)}\left(\pm 4\Omega \sqrt{m_{0}r_0} \right)}\approx -2\, m_0 \pi\epsilon r r_0 \Omega^2 ,
\end{equation}
and hence
\begin{equation}
\label{o10b}
\lim_{\Omega \to 0} m(r, \omega)\approx \left.-2\, m_0 \pi\epsilon r r_0 \Omega^2\right|_{\omega=\pm\Omega}.
\end{equation}
We compare this with the Fourier transform of a phenomenological retarded profile,
\begin{equation}
\label{o8}
F(r, t) = -i m_0 \big[ 1 + \epsilon \cos\left[\Omega (t - r)\right] \big].
\end{equation}
For non-zero $\omega$,
\begin{equation}
\label{o9}
F(r,\omega)=-\left. im_0 \pi \epsilon\, e^{- i r \Omega}\right|_{\omega=\Omega} -\left.im_0 \pi \epsilon\,e^{i r \Omega} \right|_{\omega=- \Omega},
\end{equation}
and expanding for small $\Omega$ gives
\begin{equation}
\label{o11}
\lim_{\Omega \to 0} F(r,\omega)\approx -\left.m_0\pi\epsilon r\,\Omega \right|_{\omega=\pm\Omega}.
\end{equation}
Matching the radial dependence of Eqs.~\eqref{o10b} and \eqref{o11} fixes
\begin{equation}
\label{o12}
r_0 = \frac{1}{2\Omega}.
\end{equation}
Substituting $r_0$ into Eq.~\eqref{4} specifies the weak-field metric for an oscillating-mass black hole.

\subsection{Power from the Green's function method}
\label{oscillating_green}

We now compute the electromagnetic perturbation generated by the oscillating source using a Green's function method. With $m(r,\omega)$ and $r_0$ given by \eqref{o7} and \eqref{o12}, respectively, the effective source term follows from \eqref{28b}:
\begin{equation}
\label{g1}
S^{\phi}(\omega)=-2B_0m'(r,\omega)\sin\theta,
\end{equation}
where
\begin{equation}
\label{g2}
m'(r,\omega)=\left.-2m_0\pi\epsilon\,
\sqrt{\frac{\Omega}{r}}\,
\frac{H^{(1)}_{1}\!\left(2\sqrt{r\Omega}\right)}
     {H^{(1)}_{0}\!\left(2\sqrt{2 m_0\Omega}\right)}
\right|_{\omega=\pm\Omega}.
\end{equation}

From \eqref{P6}, the perturbation obeys
\begin{equation}
\label{o13b}
\partial_\mu\!\left(\sqrt{-g}\,g^{\mu\alpha}g^{\phi\beta}\,\delta F_{\alpha\beta}\right)=S^\phi(\omega).
\end{equation}
For notational simplicity, we write $\delta A_\mu\equiv A_\mu$ in this appendix. In the asymptotic region where the radiation is observed, the differential operator on the left-hand side is approximated by its Minkowski-space form, while the source retains the curved-spacetime information through $m'(r,\omega)$. The dominant contribution comes from the radiation-formation region $r\ll r_0$; for $r\gtrsim r_0$, rapidly varying phases suppress the net contribution by destructive interference.

Using
\begin{equation*}
A_\phi(r,\omega)=f(r)\sin^2\theta,
\end{equation*}
Eq.~\eqref{o13b} reduces to
\begin{equation}
\label{g3}
f''(r)+\left(\omega^2-\frac{2}{r^2}\right)f(r)=-2 B_0 m'(r,\omega).
\end{equation}
The homogeneous solutions are
\begin{equation*}
  f_{1}(r)= r\,h^{(1)}_{1}(r \omega ) 
\qquad \text{and} \qquad 
f_{2}(r)= r\,h^{(2)}_{1}(r \omega ),
\end{equation*}
where $h^{(1)}_{1}$ and $h^{(2)}_{1}$ are spherical Hankel functions. We require $A^\phi(r,\omega)=f(r)\sin\theta/r$ to be regular at $r=2m_0$ and outgoing as $r\to\infty$, so
\begin{equation}
\label{g4}
 f_{\text{in}}(2 m_0)=0 
\qquad \text{and} \qquad
 f_{\text{out}}\sim e^{i\omega r} \quad \text{as } r \to \infty .
\end{equation}
The required homogeneous combinations are
\begin{equation}
 \label{g5}
 f_{\text{out}}(r) \equiv f_1(r) = r\,h^{(1)}_{1}(\omega r),
\end{equation}
and
\begin{equation}
 \label{g6}
f_{\text{in}}(r) \equiv f_1(r)\,f_2(2m_0)
- f_{2}(r)\,f_{1}(2m_0).
\end{equation}
The boundary-value Green's function is
\begin{equation}
\label{g7}
G(r,r') =
\begin{cases}
\dfrac{f_{\text{in}}(r)\,f_{\text{out}}(r')}{W} 
& 2m_0 \le r \le r', \\[8pt]
\dfrac{f_{\text{in}}(r')\,f_{\text{out}}(r)}{W}, 
& r' \le r < \infty ,
\end{cases}
\end{equation}
with Wronskian
\begin{equation}
\label{g8}
W=\frac{2 i}{\omega}f_{\text{out}}(2m_0)= \frac{4 i m_0}{\omega}\,h^{(1)}_{1}\!\left(2m_0\omega\right).
\end{equation}
Thus,
\begin{equation}
f(r)=-2B_0\int_{2m_0}^{\infty} G(r,r')\,m'(r',\omega)\,dr',
\label{g9}
\end{equation}
and
\begin{equation}
\begin{aligned}
A^{\phi}(r,\omega)
&=f(r)\,\frac{\sin\theta}{r} \\
&=-\frac{2B_0\sin\theta}{r}
  \int_{2m_0}^{\infty} G(r,r')\,m'(r',\omega)\,dr' \\
&=\frac{iB_0\omega\sin\theta}{r\,f_{\text{out}}(2m_0)}
  \int_{2m_0}^{\infty} f_{\text{in}}(r_<)\,f_{\text{out}}(r_>)\\
&\quad\times m'(r',\omega)\,dr' ,
\end{aligned}
\label{g10}
\end{equation}
where $r_< = \min(r,r')$ and $r_> = \max(r,r')$.

In the far-field limit,
\begin{equation*}
    f_{\text{out}}(r)=r\,h^{(1)}_{1}(\omega r)\sim-\frac{e^{i\omega r}}{\omega},
\end{equation*}
so
\begin{equation}
\lim_{r\to\infty} A_{\phi}(r,\omega)
\;\simeq\;
-\frac{i B_0 \,
e^{i\omega r} \sin\theta}{r\, \omega}\,
\left[\frac{\omega\,J(\omega)}{f_{\text{out}}(2 m_0)}\right],
\label{g11}
\end{equation}
where
\begin{equation}
\label{g12}
J(\omega)\equiv
\int_{2 m_0}^{\infty}
f_{\text{in}}(r')\,m'(r',\omega)\,dr',
\end{equation}
and
\begin{equation}
\label{g13}
f_{\text{out}}(2 m_0)=2 m_0\, h_1^{(1)}(2 m_0\omega).
\end{equation}
Defining $\mathcal{A}(\omega)=\omega J(\omega)/f_{\text{out}}(2m_0)$, we obtain
\begin{equation}
\mathcal{A}(\omega)
=\omega\left(
I_{\text{in}}
-\frac{h^{(2)}_{1}(2m_0\omega)}{h^{(1)}_{1}(2m_0\omega)}\,I_{\text{out}}
\right),
\label{g14}
\end{equation}
with
\begin{equation}
\begin{split}
I_{\text{in}}=\int_{2m_0}^{\infty} r'\,h^{(2)}_{1}( r'\omega)\,m'(r',\omega)\,dr',\\
I_{\text{out}}=\int_{2m_0}^{\infty} r'\,h^{(1)}_{1}(r'\omega)\,m'(r',\omega)\,dr'.
\end{split}
\label{g15}
\end{equation}
Temporarily dropping the delta functions in $m'(r,\omega)$, which are restored at the end, and changing variables according to
\begin{equation*}
   z=2 \sqrt{r' \Omega} \qquad\text{and}\qquad z_0=2\sqrt{2m_0\Omega},
\end{equation*}
gives
\begin{equation}
\begin{split}
I_{\text{in}}
= -2m_0 \pi \epsilon
\int_{z_0}^{\infty}
\left(\frac{z^2}{4\Omega}\right)
h^{(2)}_{1}\!\left(\frac{\omega z^2}{4\Omega}\right)
\Re\!\left(\frac{H^{(1)}_{1}(z)}{H^{(1)}_{0}(z_0)}\right)
\,dz, \\
I_{\text{out}}
= -2m_0 \pi \epsilon
\int_{z_0}^{\infty}
\left(\frac{z^2}{4\Omega}\right)
h^{(1)}_{1}\!\left(\frac{\omega z^2}{4\Omega}\right)
\Re\!\left(\frac{H^{(1)}_{1}(z)}{H^{(1)}_{0}(z_0)}\right)
\,dz .
\end{split}
\label{g16}
\end{equation}

In the slow-oscillation limit $\Omega\to0$, $z_0\to0$, and
\begin{equation*}
\frac{1}{H^{(1)}_{0}(z_0)}
\;\simeq\;
-\,i\,\frac{\pi}{2}\,\frac{1}{L},
\end{equation*}
where
\begin{equation*}
L \equiv \gamma + \ln\!\left(\frac{z_0}{2}\right)=\gamma +\frac{1}{2}\, \ln\!\left(2 m_0\Omega\right).
\end{equation*}
Then
\begin{equation}
\begin{split}
I_{\text{in}}
= -\frac{m_0 \pi^2 \epsilon}{L}
\int_{z_0}^{\infty}
\left(\frac{z^2}{4\Omega}\right)
h^{(2)}_{1}\!\left(\frac{\omega z^2}{4\Omega}\right)
Y_1(z)
\,dz, \\
I_{\text{out}}
= -\frac{m_0 \pi^2 \epsilon}{L}
\int_{z_0}^{\infty}
\left(\frac{z^2}{4\Omega}\right)
h^{(1)}_{1}\!\left(\frac{\omega z^2}{4\Omega}\right)
Y_1(z)
\,dz .
\end{split}
\label{g17}
\end{equation}
The integrands are rapidly oscillatory except near the endpoint region where $z=O(\sqrt{\Omega})$. Setting
\begin{equation*}
 y=\frac{z}{\sqrt{\Omega}} \quad\text{and}\quad y_0=\frac{z_0}{\sqrt{\Omega}}=2\sqrt{2 m_0},
\end{equation*}
and using
\begin{equation*}
  Y_1(z) \sim -\frac{2}{\pi z}=-\frac{2}{\pi y\sqrt{\Omega}},
\end{equation*}
we get
\begin{equation}
\begin{split}
I_{\text{in}}
= \frac{m_0 \pi \epsilon}{2 L}
\int_{y_0}^{\infty}
y\,
h^{(2)}_{1}\!\left(\frac{\omega\,  y^2}{4}\right)
\,dy, \\
I_{\text{out}}
= \frac{m_0 \pi \epsilon}{2 L}
\int_{y_0}^{\infty}
y\,
h^{(1)}_{1}\!\left(\frac{\omega\,  y^2}{4}\right)
\,dy .
\end{split}
\label{g18}
\end{equation}
Finally, with $u=\omega y^2/4$ and $u_0=2m_0\omega$,
\begin{equation}
\begin{split}
I_{\text{in}}
= \frac{m_0 \pi \epsilon}{\omega L}
\int_{2 m_0 \omega}^{\infty}
h^{(2)}_{1}\!\left(u\right)
\,du=\frac{m_0 \pi \epsilon}{\omega L}\,h_0^{(2)}\left(2 m_0 \omega\right), \\
I_{\text{out}}
= \frac{m_0 \pi \epsilon}{\omega L}
\int_{2 m_0 \omega}^{\infty}
h^{(1)}_{1}\!\left(u\right)
\,du=\frac{m_0 \pi \epsilon}{\omega L}\,h_0^{(1)}\left(2 m_0 \omega\right),
\end{split}
\label{g19}
\end{equation}
where the standard radiation boundary condition drops the oscillatory $h_0^{(1,2)}(\infty)$ endpoint. Substituting \eqref{g19} into \eqref{g14} gives
\begin{equation}
\label{g20}
\mathcal{A}(\omega)=
\frac{m_0 \pi \epsilon}{L}
\left[
h^{(2)}_{0}(2m_0\omega)
-\frac{h^{(2)}_{1}(2m_0\omega)}{h^{(1)}_{1}(2m_0\omega)}
\,h^{(1)}_{0}(2m_0\omega)
\right].
\end{equation}
For $m_0\ll1$, the bracketed factor approaches $2$. Thus, in the small-$\Omega$ and small-$m_0$ limits,
\begin{equation}
\label{g21}   
\lim_{r\to\infty} A^{\phi}(r,\omega)
\;\simeq\;\left.
-\frac{4i\, B_0 m_0 \pi\epsilon \,e^{i\omega r}\sin\theta}{r\, \omega \ln(2 m_0 \Omega)}\right|_{\omega=\pm\Omega}.
\end{equation}
The far-field electric and magnetic fields are
\begin{equation}
\label{g22}
B_\theta(\omega)=E_\phi(\omega)=\left.-\frac{4\, B_0m_0\pi\epsilon\, e^{i r \omega}\sin\theta}{r\ln(2 m_0\Omega)}\right|_{\omega=\pm\Omega}.
\end{equation}
The Poynting vector $\vec{S}(\omega)=\frac{1}{4 \pi}\vec{E}(\omega) \times \vec{B}(\omega)^*$ yields
\begin{equation}
\label{o21b}
S_r(\omega)=\frac{\pi B_0^2\, m_0^2\, \epsilon^2 \sin^2 \theta}{r^2\ln^2(2m_0\Omega) }\,\left[\delta(\omega-\Omega)+\delta(\omega+\Omega)\right]^2.
\end{equation}
The spectral energy density is
\begin{equation}
\label{senergydensity}
\begin{aligned}
\frac{d\mathcal{E}_B}{d\omega}
&=\frac{1}{2\pi}\int d\Omega\,\boldsymbol{S(\omega)}\boldsymbol{.}
  r^2\,\boldsymbol{\hat{r}} \\
&=\frac{4\pi B_0^2 m_0^2 \epsilon^2}{3\ln^2(2m_0\Omega)}
  \left[\delta(\omega-\Omega)+\delta(\omega+\Omega)\right]^2 .
\end{aligned}
\end{equation}
Using Parseval's theorem for monochromatic radiation,
\[
[\delta(\omega-\Omega)+\delta(\omega+\Omega)]^2
=
\frac{T_{\rm obs}}{2\pi}[\delta(\omega-\Omega)+\delta(\omega+\Omega)],
\]
where $T_{\rm obs}$ is the total observation time, the spectral power becomes
\begin{equation}
\label{spectralpowerosc}
\frac{dP_B}{d\omega}=\left.\frac{2\,B_0^2\, m_0^2\, \epsilon^2 }{3\ln^2(2m_0\Omega) }\right|_{\omega=\pm\Omega}
\end{equation}
The total emitted power is therefore
\begin{equation}
\label{o22}
P_B=\int \frac{dP_B}{d\omega}\,d\omega= \frac{4\,B_0^2\, m_0^2\, \epsilon^2 }{3\ln^2(2m_0\Omega) }.
\end{equation}
Restoring CGS units with $R_{Sch}=2Gm_0/c^2$ gives
\begin{equation}
\label{o23}
P_B\approx0.3\,\epsilon^2\, \frac{ B_0^2\,R_{Sch}^2c}{\ln^2\left(\frac{R_{Sch}\Omega}{c}\right)}.
\end{equation}

Equation \eqref{o23} shows that the oscillating source radiates monochromatically at the driving frequency $\Omega$, with a power suppressed by the factor $\ln^{-2}(2m_0\Omega)$. This logarithmic dependence originates from the small-argument behavior of the Hankel functions in the low-frequency limit $\Omega m_0\ll1$. In this regime, the source is controlled by the long-wavelength near-zone region, while the outgoing radiation is measured in the far zone. The connection between these regions introduces the logarithmic factor through $H_0^{(1)}(z_0)$, with $z_0=2\sqrt{2m_0\Omega}$. Similar logarithmic structures arise in low-frequency wave scattering and absorption by black holes, where the analytic behavior is governed by the relation between near-horizon and far-zone solutions \cite{Unruh1976,Hui2021,Saketh,Aminov2024}. Thus, as $\Omega$ decreases, the emitted power is suppressed only through a slowly varying logarithmic factor, reflecting the scale-sensitive nature of the long-wavelength response.

\subsection{Small-\texorpdfstring{$r$}{r} and small-\texorpdfstring{$\Omega$}{Omega} expansion of the source}
\label{oscillating_source_expansion}

The same result can be obtained by first expanding the source in the radiation-formation region. This provides a useful check on the Green's function calculation above. The source is still given by \eqref{g1}, with $m'(r,\omega)$ specified by \eqref{g2}. In the combined limits $r\ll r_0$ and $\Omega\to0$, the source reduces to
\begin{equation}
\label{o15}
S^{\phi}(\omega)=-\frac{4}{\sqrt{-g}} \left.\frac{ B_0m_0\pi\epsilon\sin\theta}{r \ln(2 m_0\Omega)}\right|_{\omega=\pm\Omega}.
\end{equation}
Using the same ansatz $A_\phi(\omega)=f(r,\omega)\sin^2\theta$, Eq.~\eqref{g3} becomes
\begin{equation}
\label{maxwellfourier}
f''(r)+\left(\omega^2-\frac{2}{r^2}\right)f(r)=-4\,\left.\frac{ B_0m_0\pi\epsilon}{r \ln(2 m_0\Omega)}\right|_{\omega=\pm\Omega}.
\end{equation}
The solution may be written as
\begin{equation}
\label{o16}
\begin{aligned}
A^\phi(\omega)
&=\sqrt{\frac{2}{\pi}}\Bigg[
\frac{(r\omega c_1+c_2)\cos(r\omega)}{r\,\omega^{3/2}}\\
&\qquad
+\frac{(-c_1+r\omega c_2)\sin(r\omega)}{r\,\omega^{3/2}}
\Bigg]\frac{\sin\theta}{r}\\
&\quad + f_p(r,\omega)\frac{\sin\theta}{r},
\end{aligned}
\end{equation}
with particular solution
\begin{equation}
\label{o17}
f_p(r,\omega)=\left.-\frac{4\, B_0 m_0 \pi \epsilon}{r\, \omega^2 \ln(2 m_0\Omega)}\right|_{\omega=\pm\Omega}.
\end{equation}
Regularity at the inner boundary and outgoing behavior at infinity give
\begin{equation}
\label{o18}
c_2= \left.2\sqrt{\frac{2 \pi}{\omega}}\frac{B_0m_0\epsilon}{\ln(2 m_0\Omega)}\right|_{\omega=\pm\Omega},
\end{equation}
and
\begin{equation}
\label{o19}
c_1=-\left.2i\sqrt{\frac{2 \pi}{\omega}}\frac{B_0m_0\epsilon}{\ln(2 m_0\Omega)}\right|_{\omega=\pm\Omega}.
\end{equation}
Substituting these into \eqref{o16}, the asymptotic fields become
\begin{equation}
\label{o20}
B_\theta(\omega)=E_\phi(\omega)=\left.-\frac{4\, B_0m_0\pi\epsilon\, e^{i r \omega}\sin\theta}{r\ln(2 m_0\Omega)}\right|_{\omega=\pm\Omega}.
\end{equation}
Thus,
\begin{equation}
\label{o21c}
S_r(\omega)=\frac{\pi B_0^2\, m_0^2\, \epsilon^2 \sin^2 \theta}{r^2\ln^2(2m_0\Omega) }\,\left[\delta(\omega-\Omega)+\delta(\omega+\Omega)\right]^2.
\end{equation}
Integrating over angles gives the spectral energy
\begin{equation}
\label{senergydensityb}
\frac{d\mathcal{E}_B}{d\omega}
=\frac{4\pi B_0^2 m_0^2 \epsilon^2}{3\ln^2(2m_0\Omega)}
  \left[\delta(\omega-\Omega)+\delta(\omega+\Omega)\right]^2 .
\end{equation}
Using Parseval's theorem,
\[
[\delta(\omega-\Omega)+\delta(\omega+\Omega)]^2
=
\frac{T_{obs}}{2\pi}[\delta(\omega-\Omega)+\delta(\omega+\Omega)],
\]
we obtain the spectral power
\begin{equation}
\label{spectralpoweroscb}
\frac{dP_B}{d\omega}=\left.\frac{2 B_0^2\, m_0^2\, \epsilon^2 }{3\ln^2(2m_0\Omega) }\right|_{\omega=\pm\Omega}
\end{equation}
and finally, the total power becomes
\begin{equation}
\label{o22b}
P_B
=\int \frac{dP_B}{d\omega}\,d\omega=
\frac{4 B_0^2 m_0^2 \epsilon^2}{3\ln^2(2m_0\Omega)} .
\end{equation}
Restoring CGS units gives
\begin{equation}
\label{o23b}
P_B\approx0.3\, \epsilon^2\, \frac{ B_0^2\,R_{Sch}^2c}{\ln^2(\frac{R_{Sch}\Omega}{c})}.
\end{equation}
This agrees with \eqref{spectralpowerosc} and \eqref{o23}. Hence the Green's function method and the direct near-zone source expansion give the same monochromatic power, confirming that the radiation is controlled by the small-$r$, low-frequency structure of the source.

\section{Trial Case II: Black hole with exponentially decaying mass}
\label{Laplace}

As a second auxiliary example, we consider a central mass that decays exponentially. This appendix is not meant to rederive the causal Laplace construction developed in Sec.~\ref{Hawking}; instead, it shows how the same boundary-value machinery behaves for a simpler decay law. Thus the stationary past, one-sided Laplace transform, bounded $K_0$ branch, and outgoing analytic continuation are inherited directly from Eqs.~\eqref{Hmassprofile}--\eqref{Houtgoingsolution}.

The prescribed boundary mass is
\begin{equation}
\label{E1}
M(t)=m_0\, e^{-t/\tau_{\rm H}},
\end{equation}
where $m_0$ is the initial mass and $\tau_{\rm H}$ is a large characteristic decay time. Evaluating the boundary condition at $r=2m_0$ gives
\begin{equation}
\label{E2}
m (r=2m_0,s)=\frac{m_0}{s\,\left(1+s\tau_{\rm H}\right)} .
\end{equation}
Inserting this boundary data into the causal outgoing solution \eqref{Houtgoingsolution} determines
\begin{equation}
\label{E3}
c_1(s)=\frac{ m_{0} }
     {  s\bigl(1 + s \tau_{\rm H} \bigr)\,
       H_{0}^{(1)}\left( 4i \sqrt{m_{0}r_0} \, s \right) } .
\end{equation}
Therefore the exponential trial profile is
\begin{equation}
\label{E4}
m(r,s)=
\frac{m_0 H_0^{(1)}\!\left(2i\sqrt{2rr_0}\,s\right)}
{s\left(1+s\tau_{\rm H}\right)
H_0^{(1)}\!\left(4i\sqrt{m_0r_0}\,s\right)} .
\end{equation}
The small-$s$ expansion separates the local decay law from the first nontrivial radial correction:
\begin{equation}
\label{E6}
m(r,s)\approx m_0\left(\frac{1}{s}-\tau_{\rm H}\right)+2\, m_0\,r\,r_0\,s .
\end{equation}
To fix the propagation scale, we compare this radial correction with the retarded exponential profile $\mathcal{L}(r,t)=-m_0\,e^{-(t-r)/\tau_{\rm H}}$. Its Laplace transform is
\begin{equation}
\label{E7}
\begin{aligned}
\mathcal{L}(r,s)
&=-\int_0^{\infty} e^{-s t}
\left(m_0 e^{-(t-r)/\tau_{\rm H}}\right)\,dt \\
&=-\frac{e^{r/\tau_{\rm H}}m_0\tau_{\rm H}}{1+s\tau_{\rm H}} .
\end{aligned}
\end{equation}
To first order in $r$ and in the slow-decay limit, this gives
\begin{equation}
\label{E8}
\mathcal{L}(r,s)\approx -m_0\,r\,\left(1-s\tau_{\rm H}\right).
\end{equation}
Matching the $\mathcal{O}(rs)$ terms in Eqs.~\eqref{E6} and \eqref{E8} gives the same scale relation as in the Hawking calculation:
\begin{equation}
\label{E9}
r_0=\frac{\tau_{\rm H}}{2} .
\end{equation}
This confirms that the length scale $r_0$ is controlled by the causal propagation time of the mass-loss profile, not by the specific functional form of the decay.

Substituting $r_0=\tau_{\rm H}/2$ into \eqref{E4}, the near-zone source is controlled by the small-argument limit of the Hankel functions. In the same low-frequency regime used in Sec.~\ref{Hawking}, the leading radial derivative is
\begin{equation}
 \label{Eii} 
 m'(r,s)=m_0\left(\frac{1-s\tau_{\rm H}}{rs}\right) .
\end{equation}
Substituting this into \eqref{28b}, the source term in Laplace space becomes
\begin{equation}
\label{E13}
S^{\phi}(s)=-\frac{2\,B_0 m_0 \sin\theta}{r} \left( \frac{1 - s \tau_{\rm H}}{s}\right) .
\end{equation}
Solving the same asymptotically flat Maxwell problem as in Sec.~\ref{Hawking}, with the ansatz $A_\phi(s)=f(r,s)\sin^2\theta$ and imposing regularity at the inner boundary together with a decaying retarded field at infinity, gives the far-zone vector potential
\begin{equation}
\label{VecPot}   
A^{\phi}(s)\simeq - \frac{2\,B_0\, m_0\, e^{-r s} \sin\theta}{r}\left(\frac{1-s\tau_{\rm H}}{s^2}\right) .
\end{equation}
Thus, in the combined limits $r\to\infty$ and $|s|\to0$, the leading electric and magnetic fields are
\begin{equation}
\label{E15}
B_\theta(s)=E_\phi(s)\simeq- \frac{2\,B_0\, m_0\, e^{-r s} \sin\theta}{r\,s }+\mathcal{O}(s) .
\end{equation}
After the inverse Laplace transform and the subsequent Fourier transform, the leading nonzero-frequency fields are
\begin{equation}
\label{E16}
B_\theta(\omega)=E_\phi(\omega)\approx -\frac{2i\, B_0\,m_0\, e^{i r \omega}  \sin\theta}{r\,\omega} .
\end{equation}
The corresponding spectral energy is
\begin{equation}
\label{exponentialspectralenergy}
\frac{d \mathcal{E}_B}{d\omega}\approx\frac{  2\,B_0^2\,m_0^2}{3 \pi^2\omega^{2}} .
\end{equation}
This expression describes the spectral energy associated with the full exponential decay history. The appropriate infrared cutoff is therefore set by the duration of the decay, 
\begin{equation}
\omega_{\min}\sim \frac{1}{\tau_{\rm H}},
\end{equation}
while the ultraviolet cutoff is again taken to be $\omega_{\max}\sim c/R_{Sch}$. Integrating over the emitted frequency gives
\begin{equation}
\label{exponentialtotalenergy}
\begin{aligned}
\mathcal{E}_B
&=\int^{\omega_{max}}_{\omega_{min}}\frac{d\mathcal{E}_B}{d\omega}\,d\omega \\
&=\frac{2}{3\pi^2}B_0^2m_0^2
\left(\frac{1}{\omega_{min}}-\frac{1}{\omega_{max}}\right) .
\end{aligned}
\end{equation}
The lifetime-averaged emitted power is obtained by dividing by the decay timescale $\tau_{\rm H}$:
\begin{equation}
L_B=\frac{\mathcal{E}_B}{\tau_{\rm H}}
=
\frac{2}{3 \pi^2}\frac{B_0^2m_0^2}{\tau_{\rm H}}
\left(\frac{1}{\omega_{min}}-\frac{1}{\omega_{max}}\right) .
\end{equation}
Taking $\omega_{min}=1/\tau_{\rm H}$, $\omega_{max}=c/R_{Sch}$, and $R_{Sch}=2\,Gm_0/c^2$, the average emitted power in CGS units becomes
\begin{equation}
L_B\approx0.02\:B_0^2\,R_{Sch}^2\,c\left[1-\frac{R_{Sch}}{c\tau_{\rm H}}\right] .
\end{equation}

For an observation time shorter than the decay timescale, $T_{obs}<\tau_{\rm H}$, one should not obtain the power by simply replacing $\omega_{\min}$ with $1/T_{obs}$ in the full-history spectrum above. In that case the observed spectrum depends on the finite time window, and a proper calculation would require the Fourier transform of the truncated time-domain field over the interval $T_{obs}$. Since this exponential profile is used only as an auxiliary consistency check, we restrict the quoted power to the decay-timescale-averaged result.

\section{Problem with the Vaidya metric}
\label{Vaidya}

We first consider the Vaidya metric, which describes a spherically symmetric spacetime with radial null energy flux. In outgoing Eddington-Finkelstein coordinates $(u,r,\theta,\phi)$, the line element is
\begin{equation}
ds^2=-\left(1-\frac{2M(u)}{r}\right)du^2-2\,du\,dr+r^2d\Omega^2 .
\label{V1}
\end{equation}
In the static limit, $M(u)=M_0$, this reduces to the Schwarzschild geometry written in outgoing Eddington-Finkelstein coordinates.

To evaluate the effective source term, we use the same static magnetic-field configuration as in the Schwarzschild case:
\begin{equation*}
A_\phi^{(0)}=\frac{B_0}{2}\,r^2\sin^2\theta ,
\label{V2}
\end{equation*}
which gives
\begin{equation}
F^{(0)}_{r\phi}=B_0\,r\sin^2\theta,
\qquad
F^{(0)}_{\theta\phi}=B_0\,r^2\sin\theta\cos\theta .
\label{V3}
\end{equation}
For the metric \eqref{V1}, the relevant inverse components are
\begin{equation}
\begin{split}
&g^{ur}=g^{ru}=-1,\\
&g^{rr}=1-\frac{2M(u)}{r},\\
&g^{\theta\theta}=\frac{1}{r^2},\\
&g^{\phi\phi}=\frac{1}{r^2\sin^2\theta}.
\label{V4}
\end{split}
\end{equation}
Thus, the effective source term from \eqref{P7} is
\begin{equation}
\begin{aligned}
S^\phi
&=\partial_\mu\!\left(\sqrt{-g}\,F^{(0)\mu\phi}\right)\\
&=\partial_u\!\left(\sqrt{-g}\,F^{(0)u\phi}\right)
 +\partial_r\!\left(\sqrt{-g}\,F^{(0)r\phi}\right)\\
&\quad
 +\partial_\theta\!\left(\sqrt{-g}\,F^{(0)\theta\phi}\right) .
\end{aligned}
\label{V6}
\end{equation}
Using \eqref{V3} and \eqref{V4}, we obtain
\begin{equation}
S^\phi=0.
\label{V9}
\end{equation}
Thus, even though the mass function $M(u)$ introduces time dependence through a null coordinate, the effective electromagnetic source vanishes identically. This reflects the fact that the Vaidya solution describes a highly symmetric flux of classical null radiation, which does not generate the spatial inhomogeneity required to produce a nontrivial electromagnetic source in the present framework.

\section{Problem with Oppenheimer--Snyder model}
\label{Oppenheimer}

We next consider the Oppenheimer--Snyder model of gravitational collapse. Although this model describes the collapse of a massive object leading to black hole formation, we examine its structure here to assess whether its time-dependent boundary can generate an electromagnetic source in the present framework.

The spacetime is divided into interior and exterior regions separated by a stellar boundary $R_b$. We first recall the standard OS form, and then introduce the time dependence relevant to the present discussion through the boundary, $R_b(\tau)=2M(\tau)$. For the exterior region $R>R_b$, the geometry is described by the Schwarzschild metric:
\begin{equation}
\label{Sch}
ds^2=-\left(1-\frac{r_0}{r}\right)dt^2+\frac{dr^2}{\left(1-\frac{r_0}{r}\right)}+r^2d\Omega^2,
\end{equation}
where $r_0=2 M_0$.

For $R \leq R_b$, the interior metric is
\begin{equation}
\label{OS1} 
ds^2=-d\tau^2+e^{\bar{\omega}}dR^2+e^{\omega}d\Omega^2
\end{equation}
in co-moving coordinates $(\tau,R,\theta,\phi)$, which are related to Schwarzschild coordinates as discussed in \cite{OS1}-\cite{OS2}. Solving the field equations for \eqref{OS1} yields
\begin{equation*}
 \begin{split}
 &e^\omega=r^2=R^{2} \left( 1 - \frac{3 \sqrt{r_{0}} \, \tau}{2 R_{b}^{3/2}} \right)^{4/3},\\
 &e^{\bar{\omega}}=\left( 1 - \frac{3 \sqrt{r_{0}} \, \tau}{2 R_{b}^{3/2}} \right)^{4/3},
 \end{split}    
 \end{equation*}
and the resulting interior line element becomes
\begin{equation}
\label{OS2}
ds^2=-d\tau^2+
\left( 1 - \frac{3 \sqrt{r_{0}} \, \tau}{2 R_{b}^{3/2}} \right)^{4/3}
\left(dR^2+R^2 d\Omega^2\right).
\end{equation}

The relevant time dependence in the Oppenheimer--Snyder construction resides in the interior geometry and in the evolving boundary $R_b(\tau)$. The exterior region is Schwarzschild and therefore does not contain the explicit time-dependent radial structure required to source the electromagnetic perturbation. We therefore evaluate the induced source using the interior metric \eqref{OS2}, and then allow the boundary to evolve through $R_b(\tau)=2M(\tau)$.

We now consider the sourceless form of the modified Maxwell equations in this background. Substituting the metric \eqref{OS2} into static Maxwell's equations and adopting the ansatz $A_\phi(R,\theta)=f(R)\sin^2\theta$, the solution for the vector potential is
\begin{equation}
\label{OS3}
A_\phi(R,\theta)=R^2\sin^2\theta.
\end{equation}
We now allow the boundary to evolve through $R_b(\tau)=2\,M(\tau)$ and compute the induced source term from \eqref{27}. Despite this time dependence, the symmetry of the OS construction gives
\begin{equation}
\label{OS4}
 S^\phi=\frac{1}{\sqrt{-g}}\bigg[\partial_R\left(g^{RR}g^{\phi \phi}F_{R \phi}\right)+\partial_\theta\left(g^{\theta\theta}g^{\phi \phi}F_{\theta \phi}\right)\bigg]=0.
\end{equation}
Thus, even after introducing time dependence through the boundary, the Oppenheimer--Snyder model yields a vanishing source term. This behavior is analogous to the Vaidya case and reflects the highly symmetric, homogeneous nature of the interior spacetime, which prevents the generation of a nontrivial electromagnetic source.

\bibliographystyle{unsrt}  
\bibliography{references}

\end{document}